\documentclass[12pt]{elsart}
\usepackage{graphicx}
\usepackage{color}

\def\be{\begin{eqnarray}}
\def\ee{\end{eqnarray}}
\def\bc{\begin{center}}
\def\ec{\end{center}}
\def\om{\omega}

\def\prt{\partial}
\def\re{{\rm Re}}
\def\im{{\rm Im}}
\def\lsim{\stackrel{\scriptstyle <}{\phantom{}_{\sim}}}
\def\gsim{\stackrel{\scriptstyle >}{\phantom{}_{\sim}}}

\def\rmd{{\rm d}}
\begin{document}
\begin{frontmatter}
\title{Relativistic Mean-Field Model \\ with Scaled Hadron
Masses and Couplings \\ }
\author[JINR,Mold]{A.S.~Khvorostukhin},
\author[JINR,GSI]{V.D.~Toneev}
\and \author[MEPHI,GSI]{D.N. Voskresensky}
\address[JINR]{Joint Institute for Nuclear Research,
 141980 Dubna, Moscow Region, Russia}
\address[Mold]{ Institute of Applied Physics, Moldova Academy of Science,
MD-2028 Kishineu, Moldova}
\address[GSI]{GSI, Plankstra\ss{}e 1, D-64291 Darmstadt, Germany}
\address[MEPHI]{Moscow Engineering Physical Institute,\\ Kashirskoe
  Avenue 31, RU-115409 Moscow, Russia}

\begin{abstract}
Here we continue to elaborate properties of the  relativistic
mean-field based model (SHMC) proposed in ref. \cite{KTV} where
hadron masses and coupling constants depend on the $\sigma$-meson
field. The validity of approximations used in \cite{KTV} is
discussed. We additionally incorporate contribution of meson
excitations to the equations of motion.  We also estimate the
effects of the particle width.  It is demonstrated that the
inclusion of the baryon-baryon hole and baryon-antibaryon loop
terms, if performed perturbatively, destroys the consistency of
the model.

\end{abstract}\end{frontmatter}

\section{Introduction}

 In recent years there has been a great interest in the
description of hadronic properties of strongly interacting matter.
It is based on the fact that various experiments indicate
modifications of hadron masses and  widths in the medium (see for
example~\cite{M07}). As expected previously, these changes are
possibly related to a partial chiral symmetry restoration in hot
and/or dense nuclear matter, cf. ~\cite{RW}. Later on it was
realized that the connection between the chiral condensate of QCD
and hadronic spectral functions was not as direct as it was
originally envisaged. Nevertheless, the study of an in-medium
modification of hadrons  is an essential point of scientific
programs at new heavy ion facilities at FAIR
(Darmstadt)~\cite{FAIR}, NICA (Dubna)~\cite{NICA} and low-energy
campaign at RHIC (Brookhaven)~\cite{RHIC_low}.

Theoretical predictions for critical baryon density and
temperature of the hadron-quark phase transition  depend
sensitively on the Equation of State (EoS) at high densities and
temperatures. In \cite{KTV} and here we focus on the study of the
EoS of the hadronic matter. Any EoS of hadronic matter should
satisfy experimental information extracted from the description of
global characteristics of atomic nuclei such as the saturation
density, the binding energy per particle, the compressibility, the
asymmetry energy and some other. Definite constraints on hadronic
models of EoS are coming  from the analysis of  direct and
elliptic flows  in Heavy-Ion Collisions (HIC).  In addition to
these constraints astrophysical bounds on the high-density
behavior of $\beta$-equilibrium neutron star matter should be
applied, see ~\cite{Army}.

In \cite{KTV}, we constructed a phenomenological Relativistic
Mean-Field (RMF) based model that allows one to calculate particle
in-medium properties and the EoS of hadronic matter in a broad
density-temperature region. The validity of this model was
demonstrated for the description of heavy ion collisions in a
broad collision energy range. Microscopically based approaches, as
the Dirac--Brueckner--Hartree--Fock method, see \cite{Fuchs}, are
very promising but need rather involved calculations. The model of
ref. \cite{KTV} is a generalization to finite temperatures of the
RMF model developed in \cite{KV04} and applied in ~\cite{Army}
(KVOR model) for describing neutron star properties.

Following ref. \cite{KV04} we assume relevance of the  (partial)
chiral symmetry restoration at high baryon  densities and/or
temperatures ~\cite{chiral} {\em{ manifesting in the form  of
the Brown-Rho scaling hypothesis}} \cite{BR}: Masses and coupling
constants of all hadrons decrease with the density increase
approximately in the same way. In \cite{KTV} we followed the
simplest form of the scaling hypothesis and scaled the quadratic
(mass) terms of $\sigma$, $\omega$, and $\rho$ fields, as well as
the nucleon mass, by a universal scaling function $\Phi $ which
was assumed to be dependent on the $\sigma$ mean field. In order
to obtain a reasonable EoS, the meson-nucleon coupling constants
were also scaled with the $\sigma$ mean field treated as an order
parameter. Differences in  scaling functions for the effective
masses of the $\omega$- and $\rho$-fields and their couplings to a
nucleon allowed us to get an appropriate density-dependent
behavior of both the total energy and the nuclear asymmetry
energy, in agreement with the constrains obtained from  neutron
star measurements, cf. ~\cite{Army,Fuchs1}.

 Note that the idea of the dropping of the meson effective
masses continues to be " a hot point" being extensively discussing
in the literature. There exist  works which simulate different
modifications of the simplest form of the scaling trying to find
an optimal ansatz. E.g., the model \cite{LKB} introduces a common
dropping of the $\om$, $\rho$ effective masses, whereas $\sigma$
is treated differently, as purely classical field, i.e. the static
space-independent order parameter. Scalings of the $\om$, $\rho$
effective masses on the one hand and the nucleon effective mass on
the other hand  are assumed to be different. Couplings are
evaluated following quark counting. As in \cite{KTV} and in the
given paper, $\om$, $\rho$ mesons are assumed to be coupled only
to the classical $\sigma$ field, since in the quark model they are
made of a quark and an antiquark, which couple oppositely to the
vector field. A support for the common dropping of the $N, \sigma,
\omega, \rho$ masses comes from lattice QCD in the strong coupling
limit \cite{OKM08} where it was found that meson masses are
approximately proportional to the equilibrium value of the chiral
condensate.

There exist models, which  do not accept  the idea of the dropping
of the effective meson masses at all. E.g., most of the RMF models
continue to use the constant $\sigma$, $\om$, $\rho$ effective
masses.   Some models introduce field interaction terms leading to
an increase of the $\sigma$, $\om$, $\rho$ effective masses with
the increase of the nucleon density, e.g., see \cite{HP,FST}. Ref.
\cite{DEI} suggests an increase rather than a decrease of the
$\rho$ meson mass with increase of the temperature, motivating it
by mixing of vector and axial mesons at finite temperature, that
authors consider as an indication towards chiral symmetry
restoration. Another models simulate only the $\rho$ width rather
than a modification of the mass, although from general point of
view a modification of the imaginary part of the self-energy of
the resonance  should stimulate a modification of the real part of
the self-energy (effective mass), as a consequence of the
Kramers-Kronig relation.

In present paper, as in our previous paper \cite{KTV}, we avoid
discussion of these interesting theoretical questions. Instead we
will follow the Brown-Rho scaling hypothesis in its simplest form
confronting further the results of the model with the HIC data.
Besides the nucleon and meson $\sigma$, $\omega$ and $\rho$ mean
fields, we included low-lying non-strange and strange baryon
resonances, meson excitations $\sigma (600)$, $ \rho (770)$,
$\omega (782)$ constructed on the ground of mean fields, and the
(quasi)Goldstone excitations $\pi (138)$, $K(495)$, $\eta (547)$
as well as  their high mass partners in the $\rm{SU(3)}$ multiplet
$K^* (892)$, $\eta^{'} (958)$, and $\varphi (1020)$. All
corresponding antiparticles are also comprised. Interactions with
mean fields are incorporated as well. In ref.~\cite{LK02} it was
shown that it is possible to reproduce particle scattering data
when the lowest baryon octet and decouplet are assumed to be the
only relevant degrees of freedom. Therefore we do not consider
higher resonances within our model.

In order to construct a practical model in \cite{KTV}, we used
several simplifications. First, we assumed the validity of
{\em{the quasiparticle approximation}} for all baryons and mesons.
Second, we supposed that {\em{baryons and meson excitations
interact only via $\sigma$, $\omega$ and $\rho$ mean fields}}.
Thus, the fermion-fermion hole and the fermion-antifermion loop
diagrams for boson propagators and the boson-fermion loop diagrams
for fermion propagators were disregarded. Also, meson-meson
excitation interactions were neglected. Thus, effectively
excitations were considered as an ideal gas of quasiparticles.
{\em{Treating meson excitations perturbatively}} we have omitted
their contribution in the equations of motion. With this Scaled
Hadron Mass-Coupling (SHMC) model we constructed the EoS as a
function of the temperature and the baryon density and used this
EoS in a broad density-temperature region to describe properties
of hot and dense matter in heavy ion collisions.

Note that  the standard RMF models generalized to finite
temperatures have been studied in the literature, e.g., see
\cite{Wal2,Freedman,Mar}. In \cite{Wal2} temperature dependence
was included only into nucleon distributions. A general treatment
of meson excitations has been considered  within the imaginary
time formalism \cite{Freedman} and  a more convenient real time
formulation \cite{Mar}. We incorporate fluctuative terms expanding
fields near their mean-field values. Simplifying we retain only
quadratic fluctuations. Thus in our model the gas of excitations
interacts only through mean fields. Within this approximation our
results can be reproduced using above mentioned finite temperature
quantum field theory techniques.

 In the present paper, we check the validity of different approximations
assumed in \cite{KTV} and consider several possibilities how the
model can further be improved. In sect. \ref{about} we introduce
the SHMC model of \cite{KTV}. In Sect. \ref{pres}, the pressure
functional of the model is constructed and the equations of motion
are derived.  Boson excitation terms are incorporated in the
equations of motion and a comparison is made with the perturbative
treatment carried out in \cite{KTV}. Section \ref{res} estimates
the effects of finite particle widths. In Appendix A, we discuss
differences in two possible treatments of the $\sigma$ meson
field, first,  as an order parameter (as in \cite{KTV}) and
second, as an independent variable, i.e., considering $\sigma$ on
equal footing with other field variables ($\om$ and $\rho$).
Appendix B demonstrates problems which arise if  the baryon loop
terms are included. Fermion loop effects  on the boson excitation
masses are evaluated within a perturbation theory approach and arguments
are given why these effects are not included into the SHMC model.

 In reality the nucleon self-energies have a momentum
dependence which  is not so small.   It manifests itself in  high
energy heavy-ion collisions \cite{Mar1} and affects
 different  properties of atomic nuclei  \cite{Mar2}.
P-wave pion- and kaon-baryon interactions may significantly affect
properties of the pion and kaon sub-systems, see
\cite{MSTV90,V93,KV03} and refs. therein. As in \cite{KTV} and in
most of RMF models, here we continue to disregard the p-wave
effects.

A number of other important effects is not incorporated into
our model. However the full theoretical quantum field description
of many strongly interacting hadron  species can't be constructed
in any case. Using RMF based models and their generalizations one
should always balance between a realistic and practically
tractable descriptions. Thus we postpone with further
generalizations of the SHMC model. Further improvements of the
model will be done after it will pass the check   in  actual
hydrodynamical calculations of heavy ion collisions in a broad
energy regime, that is our future program.

\section{About the SHMC model}\label{about}

Following \cite{KV04} we use the $\sigma$-field dependent
effective masses of baryons
 \be \label{bar-m}
{m_b^*}/{m_b}=\Phi_b(\chi_\sigma  \sigma)= 1 -g_{\sigma b} \
\chi_{\sigma} \ \sigma /m_b \,, \; b\in\{b\}
 \ee
 with the baryon set

 $\{b\}=N(938)$, $\Delta (1232)$,
$\Lambda (1116)$, $\Sigma (1193) $, $\Xi (1318)$, $\Sigma^*
(1385)$, $\Xi^* (1530)$, and $\Omega (1672) +$ all antibaryons.

The mass terms of the mean fields are
 \be \label{bar-m1}
{m_m^*}/{m_m}&=&|\Phi_m (\chi_\sigma \sigma)|\,, \quad
m\in\{m\}=\sigma,\om,\rho\,,
 \ee
 where $g_{\sigma b}$ are the $\sigma b$-coupling constants.

For the sake of simplicity we scale all  couplings $g_{\sigma b}$
by a single scaling function $\chi_{\sigma}(\sigma )$, and all
$g_{\om b}$, $g_{\rho b}$ by $\chi_{\om}(\sigma )$ and
$\chi_{\rho}(\sigma )$ scaling functions, respectively. Therefore,
all scaling functions depend only on $\sigma $~\cite{KTV}. The
idea behind that is as follows.  The $\sigma $-field can be
interpreted as an effective field simulating a response of the
$ud$-quark condensate. The change of effective hadron masses and
couplings is associated, namely, with a modification of the quark
condensate in matter. Thus, we consider the $\sigma$-field as a
composite field, like an order parameter, whereas other meson
fields are treated as fundamental fields. The $\sigma$ excitations
are then interpreted as fluctuations around the mean value of the
order parameter. Similarly, long-scale fluctuations  are treated
in the Landau phenomenological theory of phase transitions.

To single out quasiparticles (excitations) from the mean fields,
one should do the following replacements  in the Lagrangian:
$\om_0 =\om_0^{\rm cl} +\om^{\prime} $, $R_0 =R_0^{\rm cl}
+R_0^{\; \prime}$,
 $\vec{\om}=\vec{\om}^{\; '}$ and $\vec{\rho}=\vec{\rho}^{\; '}$.
Here $\om_0^{\rm cl}$, $R_0^{\rm cl}$ are the mean (classical)
field variables and $\om_\mu^{\prime}$, $(\rho^{\prime}_0
)^{\mu}$, $(\rho^{\prime}_{\pm} ) ^{\mu}$ are responsible for  new
excitations, $R_0 =\rho_0^3$. In~\cite{KTV} we constructed a
thermodynamic potential that besides mean-field terms includes the
contribution of $\om$ and $\rho$ excitations. By varying with
respect to the fields $\om_0^{\rm cl}$, $R_0^{\rm cl}$ we obtain
equations of motion from where  the $\om_0^{\rm cl}(\sigma )$,
$R_0^{\rm cl}(\sigma )$ fields are extracted and put  back into
the thermodynamic potential. A similar procedure has been
used in a number of works, e.g. in \cite{LKB}. Then, in contrast
with \cite{LKB}, supposing $\sigma =\sigma^{\rm
cl}+\sigma^{\prime}$, we expand the thus obtained effective
potential in $\sigma^{\prime}$ up to squared terms (contribution
of $\sigma^{\prime}$ fluctuations) and, varying the thermodynamic
potential in $\sigma^{\rm cl}$, derive the equation of motion for
the resulting order parameter.

On the other hand, if the $\sigma$ field was treated on equal
footing with $\om_0$ and $R_0$, as it was done in the standard
Walecka model, we would consider all three fields as independent
variables. The comparison between two choices is performed in
Appendix A.

The dimensionless scaling functions $\Phi_b$ and $\Phi_m$, as well
as the coupling scaling functions $\chi_m$, depend on the scalar
field in the combination $\chi_\sigma(\sigma) \ \sigma$.
Therefore, we introduce the variable
 \be\label{f} f=g_{\sigma N} \ \chi_\sigma \
\sigma/m_N\,.
 \ee
 Following \cite{KV04} we assume approximate
validity of the Brown-Rho scaling ansatz in the simplest form
 \be
\label{Br-sc}\Phi =\Phi_N =\Phi_\sigma =\Phi_\om =\Phi_\rho =
 1-f ,
 \ee
using $\chi_{\sigma}=\Phi_{\sigma}$. Thereby, in terms of $\sigma$
one obtains $\Phi (\sigma ) =[1+g_{\sigma N}\sigma /m_N ]^{-1}$.
 One could partially break the scaling, if it were required
from comparison with the data.

\begin{figure}[t]
 \hspace*{16mm}
\includegraphics[width=100mm,clip]{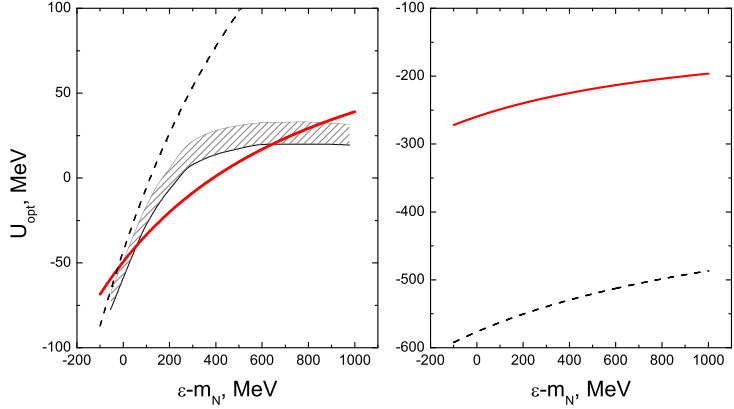}
\caption{ Energy dependence of the nucleon (left) and
antinucleon (right) optical potentials. Solid lines -- predictions
of our model and dash lines, of the original Walecka model. Shaded
area shows uncertainties in extrapolation from finite nuclei to
cold nuclear matter~\cite{FL}. }
 \label{UoptaN}
\end{figure}

We keep the standard expression for the nonlinear self-interaction
(potential $U$) of the RMF models, but now it is expressed in
terms of the new variable $f$. Using (\ref{f}) the potential $U$
can be rewritten as follows:
 \be U&=&m_N^4 (\frac{b}{3}\,f^3
+\frac{c}{4}\,f^4 ) =\frac{b m_N\,(g_{\sigma N}\,\chi_\sigma \,
\sigma)^3} {3} +\,\frac{c (g_{\sigma N}\,\chi_\sigma \, \sigma)^4}
{4}\,. \label{Unew}
 \ee
 The presence of two additional parameters, "$b$"
and "$c$", allows one to accommodate realistic values of the
nuclear compressibility and the effective nucleon mass at the
saturation density. Extra attention should be paid to the fact
that the coefficient "$c$" must be positive to deal with the
stable ground state. Values of the parameters used in our SHMC
model can be found in \cite{KTV}.
 In Fig. \ref{UoptaN} we present the  dependence of nucleon
(cf. Fig. 1 of \cite{KTV}) and antinucleon optical potentials on
the single-particle  energy. Comparison is presented with
predictions of the standard Walecka model (with only $\sigma$ and
$\omega$ mean fields). As it is seen, our model describes the
nucleon optical potential in an optional way, better than the
standard Walecka model. Differences in predictions of those models
for antinucleon optical potentials are drastic. A phenomenological
value of an antiproton optical potential is limited within the
range $-100\div -350$~MeV \cite{Mar3}, in favor of the given model
compared to the standard Walecka model. Predictions for
antiprotons are very important in a light of future experiments at
FAIR.

There are mean-field solutions of the baryon and $\sigma,\om,\rho$
meson Lagrangian $\sum_{b\in \{b\}}\mathcal{L}_b +\sum_{m\in
\{m\}}\mathcal{L}_m^{\rm MF} $~\cite{KTV}. To these terms we add
the Lagrangian density for all meson excitations
 \be \label{ex} \mathcal{L}_{\rm
ex}=\sum_{ex\in \{ex\}}\mathcal{L}_{ex}, \ \ \ \{ex\} &=&\pi^{\pm
,0}(138);K^{\pm ,0},\bar{K}^0 (495); \eta (547); \\ &&
K^{*\pm,0}(892),\eta'(958),\phi(1020);\sigma',\omega',\rho'.
\nonumber
 \ee
The set $\{g\}=(\pi, K, \eta )$ is often treated as
(quasi)Goldstone (index "$g$") bosons within the chiral SU(3)
symmetrical models. Therefore, one may not scale their masses and
couplings, as we have carried out for the mean fields $\sigma ,\om
,\rho $, cf. the set $A$ for couplings in Fig. 8 (left) of ref.
\cite{KTV} . On the other hand, one may observe, cf.
\cite{KTV,KV04}, that {\em{for the case of spatially homogeneous
system the equations for mean fields and thus their mean-field
solutions do not change if one replaces the $\sigma$, $\om_0$,
$R_0$ fields by the scaled fields $\chi_{\sigma }\sigma$,
$\chi_{\om }\om_0$ and $\chi_{\rho }R_0$, provided $\Phi_b =\Phi_m
=\chi_{m}$, and $\chi_{\rho}^{\prime}=\chi_{\rho}^2$}}
($\chi_{\rho}^{\prime}$ is the scaling function of the $\rho-\rho$
interaction $g_{\rho}$, see \cite{KTV}). If one wishes to extend
this symmetry to the case when Goldstones are included, in
addition to the scaling of masses one should scale couplings,
$g^*_{mg} =g_{mg} \ \chi_{m}$, cf. set $B$ in Fig. 8 (right)  of
\cite{KTV}.  In \cite{KTV}, we tested both possibilities $g^*_{mg}
=g_{mg}$ and $g^*_{mg} =g_{mg} \ \chi_{m}$, and referred to them
as versions without and with scaling, respectively. We  include
interaction of (quasi)Goldstones with  mean fields (for $K$ and
$\eta$, for $\pi$ it is small). As the result of this interaction,
at sufficiently large (overcritical) baryon densities there may
appear mean field solutions for (quasi)Goldstone fields signaling
of condensations of these fields. Values of critical densities are
higher for the set $B$. Since there are no experimental
indications of condensation of (quasi)Goldstone bosons in the
heavy ion collision regimes, comparing our results with
experimental data, as in \cite{KTV}, we will focus on the set $B$,
where  condensates do not occur. $K^{*\pm,0},\eta' ,\phi$ are
assumed not to couple with mean fields, since there is no
experimental information for such a coupling.

When the total Lagrangian is constructed, one can derive the
equations of motion for every field. Even for low baryon density,
the equations of motion for $\sigma$, $\om$ and $\rho$  allow
mean-field solutions $\sigma_0$, $\om_0$, $\rho^3_0$. Therefore,
we use
 \be
\sigma\equiv \sigma_0; \quad \om_{\mu}=\om_0 \ \delta_{\om 0};
\quad \rho_\mu^a = R_0 \ \delta_{a
  3} \ \delta_{\mu 0}.
  \ee
We assume that the system volume is sufficiently  large and
surface effects may be disregarded. Thus,  only spatially
homogeneous RMF solutions of  the equations of motion are
considered.

\section{Improved description of meson excitations }\label{pres}
The thermodynamic potential density $\Omega$, pressure $P$, free
 energy density $F$, energy density $E$ and entropy density $S$ are
 related as
 \be\label{presmu} &&E=F+TS, \quad
  F[f,\om_0 ,R_0 ] =\sum_{i} \mu_i n_i +\Omega\,,\quad \Omega =-P,\\
  &&\mu_i=\frac{\prt F}{\prt n_i}
\,.
  \ee
Summation index $i$ runs over all particle species; $n_i$ are
particle densities. Chemical potentials $\mu_i$ enter into the
Green functions in the standard gauge combinations $\varepsilon_i
+\mu_i$.

 Thermodynamic quantities  (\ref{presmu}) can be found
from the energy-momentum tensor $T_{\mu \nu}$ which is  defined by
our Lagrangian. The energy density $E$ and pressure $P$ are given
by the diagonal terms of this tensor
 \be \label{E-P}
E=\left< T_{00}\right>, \ \ \ \ \ \
P=\frac{1}{3}\left<T_{ii}\right>~.
 \ee
 In~\cite{KTV},  the energy was chosen as a generating functional.
Here we will use the pressure functional since it is more suitable
to treat meson excitation effects in the presence of the mean
fields and baryon-loop contributions.

\subsection{Pressure at finite density and
temperature}\label{pres1}

The  pressure can be presented as the sum of the mean $\sigma$-,
$\omega$-, $\rho$-field terms as well as of contributions of
baryons and of all meson excitations. So we have
 \be
\label{Efun} P[f,\om_0 , R_0 ] &=& \sum_{m\in\{m\}}
P^{\rm MF}_m [f, \om_0 ,R_0 ]+\sum_{b\in\{b\}} P_b [f, \om_0 ,R_0
] \nonumber\\ &+& P_{\rm bos. ex.}[f,\om_0 ,R_0 ] ~.
 \ee
The first two sums are included in every RMF model but with a
smaller set $\{b\}$, whereas the boson excitation term $P_{\rm
bos. ex.}$  is constructed in \cite{KTV} beyond the scope of the
RMF approximation and will be further elaborated here.

Although in our treatment of the $\sigma$ variable all terms in
(\ref{Efun}) are functions only of $f$ and $T$, we also present
them as functions of $\om_0$ and $R_0$ in such a way that  values
of the $\om_0 (f)$ and $R_0 (f)$ mean fields can be found by
minimization of the pressure. Then $\om_0 (f)$ and $R_0 (f)$ are
plugged back in the pressure functional that becomes a function of
$f$ only. The equilibrium value of $f$ can be found by subsequent
minimization of the resulting pressure in this field.

 In a self-consistent treatment, equations of motion for the mean
 fields render
 \be \frac{\prt}{\prt \om_0 }\,P [f,\om_0 ]=0\,~, \quad
\frac{\prt}{\prt R_0 }\,P [f,R_0 ]=0\,,\label{extreme1} \ee
 and
 \be \frac{d}{d f}\,P[f,\om_0 (f), R_0 (f)]=\frac{\partial}{\partial f}\,
 P[f,\om_0 (f), R_0 (f)]=0\, \label{extremef1}
  \ee
 with pressure $P$ given by eq. (\ref{Efun}). Since
$P_{\rm bos. ex.}[f,\om_0,R_0 ]$ depends on the mean fields, its
minimization produces extra terms in the equations of motion for
the mean fields. In differentiating in (\ref{extremef1}) we used
(\ref{extreme1}). This self-consistency of the scheme allows us to
be sure of thermodynamic consistency of the model.

In \cite{KTV},  excitations were treated perturbatively.
Accordingly, we assumed that $P_{\rm bos. ex.} =P_{\rm bos.
ex.}[f^{\rm MF},\om_0^{\rm MF} ,R_0^{\rm MF} ]$, where $f^{\rm
MF},\om_0^{\rm MF} ,R_0^{\rm MF}$ are found by minimization of the
pressure without inclusion of the boson excitation term. Thus,
equations of motion for mean fields that we used in \cite{KTV}
are:
 \be
&&\frac{\prt}{\prt \om_0 }\,\left[\sum_{m\in\{m\}} P^{\rm MF}_m
[f, \om_0 ,R_0 ]+\sum_{b\in\{b\}} P_b [f, \om_0 ,R_0 ]\right]
=0\,~, \nonumber\\ &&\frac{\prt}{\prt R_0
}\,\left[\sum_{m\in\{m\}} P^{\rm MF}_m [f, \om_0 ,R_0
]+\sum_{b\in\{b\}} P_b [f, \om_0 ,R_0 ]\right] =0\,
\label{extreme}
 \ee
 and
 \be &&\frac{d}{d f}\,\left[\sum_{m\in\{m\}}
P^{\rm MF}_m [f, \om_0 ,R_0 ]+\sum_{b\in\{b\}} P_b [f, \om_0 ,R_0
]\right] \nonumber\\&&=
 \frac{\prt\,}{\prt f }\left[\sum_{m\in\{m\}}
P^{\rm MF}_m [f, \om_0 ,R_0 ]+\sum_{b\in\{b\}} P_b [f, \om_0 ,R_0
]\right] = 0\,. \label{extremef}
 \ee
  Equation (\ref{extreme}) was used in differentiating  in (\ref{extremef}).
Below in Figs. \ref{KTV}--\ref{Trad1} we demonstrate how effects
of a nonperturbative treatment of boson excitations, incorporated
in (\ref{extreme1}) and (\ref{extremef1}) (a self-consistent
analysis) and neglected in (\ref{extreme}), (\ref{extremef}),
affect results of the SHMC model.

Actually, in \cite{KTV} instead of varying the pressure at fixed
chemical potentials $\mu_i$ and the temperature $T$, we varied the
energy density under the condition that one should not vary it
with respect to the particle occupation numbers. When $E$ is
varied, one should fix the particle densities $n_i$ and the
entropy densities $S_i$, which is equivalent to fixed  particle
occupations in our quasiparticle approach. Two procedures
mentioned are equivalent, provided baryon-baryon hole and
baryon-antibaryon excitation effects (loop contributions) are
disregarded (as we did in (\ref{extreme1}) -- (\ref{extremef})).
An attempt to incorporate the baryon loop corrections into our
scheme has been done in Appendix B.

Now let us  consider partial  contributions to the pressure in eq.
(\ref{Efun}).

\subsection{The baryon contribution}
The  contribution of the given baryon (antibaryon) species
$b\in\{b\}$ to the pressure is as follows:
 \be\label{Eb}
 P_b [f, \om_0 ,R_0 ]&=&
 \frac{1}{3}(2 s_b +1)\intop_0^{\infty}\frac{\rmd p \
p^4}{2\pi^2}\,\frac{f_b }{\om_b}-t^Q_b \ n_b \
 \mu_{\rm ch}~,  \nonumber\\
& & p=|\vec{p}|,\quad \om_b =\sqrt{m_b^{*2}(f)+p^2}.
 \ee
The spin factor $s_b =1/2$ for nucleons ($N$) and hyperons, while
$s_b =3/2$ for  $\Delta$-resonances.

The baryon set $\{b\}$  to be used   (taken from Table 1 of
\cite{KTV}) was fixed above, see after eq. (\ref{bar-m}). Little
differences in masses of charged and neutral particles of the
given species are ignored. Also we ignore small inhomogeneous
Coulomb field effects and put the electric potential $V=0$. The
charge chemical potential $\mu_{\rm ch}$ is then related to the
isospin composition of the system. For the isospin-symmetric
system,  $N=Z$, one has $\mu_{\rm ch}=0$.

The Fermi-particle (baryon/antibaryon) occupation
 \be\label{oc}
f_b &=&\frac{1}{\exp[(\om_b -\mu_b^*)/T]+1} \quad
 \ee
depends on the gauge-shifted values of the chemical potentials
 \be \label{Tmu}
 \mu_b^* =t_b  \ \mu_{{\rm bar}}+t^s_b \ \mu_{{\rm
str}}+t^Q_b \mu_{\rm ch}  -g_{\om b} \ \chi_{\om}\om_0 - t^3_b \
g_{\rho b} \ \chi_{\rho} \ R_0 ~.
 \ee
The baryon/antibaryon chemical potential of the $b$-species is
$\mu_b =t_b \ \mu_{{\rm bar}}$, and the corresponding strangeness
term is $\mu^{s}_b =t^s_b \ \mu_{{\rm str}}$. Baryon quantum
numbers $t_b$, $t_s^b$, $t_b^3$ and $t_b^Q$ are baryon charge,
strangeness, isospin projection and electric charge, respectively,
and proper charge conjugated values for antiparticles are given in
Table 1 in \cite{KTV}.

\subsection{Mean-field contribution}
It is convenient to  introduce the coupling ratios
 \be\label{rat}
 x_{mb}=g_{mb} /g_{mN}, \,\,\, m\in\{ m\}=\sigma ,\om ,\rho ,
 \ee
and, instead of $\chi_m$, other variables
 \be
\eta_{m}(f)={\Phi_m^2(f)}/{{\chi}_m^2(f)}\,, \label{eta}
 \ee
 since the pressure depends namely on this sort of combinations rather
than on $\Phi_m$ and $\chi_m$ separately.

In terms of these new variables the contribution of  mean fields
to the pressure is as follows:
 \be \label{Esig}
 P^{\rm MF}_\sigma [f]=-
\frac{m_N^4\,f^2}{2\, C_\sigma^2}\, \eta_{\sigma}(f) -{U}(f) ,
 \ee
 \be P^{\rm MF}_\om [f,\om_0 ] =
\frac{m_N^2\eta_\om(f)}{2\,C_\om^2}\left[g_{\om
N}\,{\chi}_\om\,\om_0 \right]^2\,, \label{ome}
 \ee
 \be \label{Erho}P^{\rm MF}_{\rho}[f,R_0 ]= \frac{m_N^2 \
\eta_\rho(f)}{2\,C_\rho^2}\left[g_{\rho N}\,{\chi}_\rho \ R_0
\right]^2\!  .
 \ee
Here the renormalized constants are
 \be
C_m=\frac{m_N \ g_{mN}}{m_m} . \label{Cm}
 \ee
The net baryon density is given by  \cite{KTV}:
 \be\label{bari}  n_B \equiv \sum_{b\in \{b\}}t_b n_b
 ~,\quad n_b =(2s_b +1)  \intop_0^{\infty}\frac{\rmd p \
p^2}{2\pi^2}\,f_b~,
 \ee
where $n_b$ is the  baryon (antibaryon) {\em{number}} density and
occupation baryon (antibaryon) density is defined by eq.
(\ref{oc}). On the other hand, for fixed baryon species the
contribution to the baryon density should obey the thermodynamic
consistency condition
 \be\label{cons1} n_b
=\left.\frac{\partial P}{\partial \mu_{b}^{*}}\right|_{T}.
 \ee
Both quantities presented by (\ref{cons1}) and  (\ref{bari})
coincide provided contributions of boson excitations  do not
depend on the baryon loop terms (see Appendix B). In this case
thermodynamic consistency of the model is preserved, see  below in
more detail.

The isotopic charge density in the baryon sector is given by
 \be
 n^t_{B} =2 \sum_{b\in\{b\}} \ t_b^3 \ n_b  \ x_{\rho b}~.
 \ee
 The isovector baryon density $n^t_{B}$ plays the
role of the source for the $\rho$-meson field
$\rho_0^{(3)}=R_0$\,. Therefore, for the iso-symmetrical matter
($N=Z$)  one has $n^t_B =0$ and $P^{\rm MF}_\rho =0$\,.

The net strangeness density of  baryons and mesons reads
 \be\label{nt}
 n_{\rm
 str}=\sum_{i\in\{b\},\{\rm ex\}}t_i^s \ n_i
~.
 \ee
Bearing in mind applications of the model to high-energy heavy ion
collisions from AGS to RHIC energies we assume that all strange
particles are trapped inside the fireball till the freeze-out.
Therefore, the total strangeness is zero. Thus, we put $n_{\rm
str}=0$. This condition determines the value of the strangeness
chemical potential $\mu_{\rm str}$.

Similarly, we may introduce the electric charge density
 \be n_{\rm
ch} = \sum_{i\in\{b\},\{ex\}} t_i^{Q} \ n_i ~.
 \ee
The quantity $n_{\rm ch} =(Z/A)n_{B}$ determines the value of the
charged chemical potential $\mu_{\rm ch}$. For the symmetric
matter, $N=Z$,  ignoring Coulomb effects one may put $\mu_n
=\mu_p$ and $\mu_{\rm ch}=0$.

Our SHMC model pressure functional   depends on \emph{four}
particular combinations of  functions, $\eta_{\sigma,\rho,\om}(f)$
and $U(f)$. Note that the dependence on the scaling function
$\eta_{\sigma}$ can always  be presented as part of the new
potential $U$ obtained by means of the replacement $U\to
U+\frac{m_N^4\,f^2}{2\, C_\sigma^2}\, (1-\eta_{\sigma}(f))$\,, and
vice versa, so the potential $U$ can be absorbed in the new
quantity $\eta_{\sigma}$. Thus actually only \emph{three}
independent functions enter into the pressure functional.
Equation~(\ref{Efun}) together with
eqs.~(\ref{Eb}),~(\ref{Esig}),~(\ref{ome}),~(\ref{Erho})
demonstrates explicitly the equivalence of mean-field Lagrangians
for constant fields with various parameters if they correspond to
the same functions $\eta_{\rho,\om}(f)$ and $\eta_{\sigma}$
(either $U(f)$) with the field $f$ related to the scalar field
$\sigma$ through eq.~(\ref{f}). In \cite{KTV}, we assumed
$\eta_{\sigma}=1$. Here we accept the same choice.

\subsection{Bosonic excitations}

To find the total pressure (\ref{Efun}), one should  define the
contribution of bosonic excitations. Within our model and in
agreement with~\cite{KTV} it is the sum of partial contributions
 \be \label{exden} P_{\rm bos.ex}[f,\om_0 ,R_0
,T]&=&P_{\sigma}^{\rm part}+P_{\om}^{\rm part}+P_{\rho}^{\rm
part}+P_\pi^{\rm part}\nonumber\\ &+&P_K^{\rm part}+ P_{\eta}^{\rm
part} +P_{K^*}^{\rm part} +P_{\eta^{\prime}}^{\rm part}
+P_{\phi}^{\rm part}~.
 \ee
The pressure of the pion gas is
 \be
P_{\pi}^{\rm part}&=& P_{\pi^{+}} +P_{\pi^{0}} +P_{\pi^{-}}=
\frac{1}{3}\intop_0^{\infty}\frac{\rmd p \ p^4}{2\pi^2}\,\\
&\times& \left[ \frac { f_{\pi^+}(\om_{\pi^+}(p))}{\om_{\pi^+}(p)}
+\frac{f_{\pi^0} (\om_{\pi^0}(p))}{\om_{\pi^0}(p)} +
\frac{f_{\pi^-} (\om_{\pi^-}(p))}{\om_{\pi^-}(p)}\right]~.
\nonumber
 \ee
Due to the absence of the $\om\rightarrow 2\pi$ decay the coupling
$ g_{\om \pi}^{*} =0$. For $N =Z$, the field $R_0 =0$ and
dependence of pion spectra on  $g_{\rho \pi}^{*}$ disappears.
Also, as in \cite{KTV}, we suppose $ g_{\sigma \pi}^{*} =0$
ignoring a small pion mass shift. Then for both charged and
neutral pions we may use
 \be \om_{\pi^{\pm}}(p)=\om_{\pi^0}(p)=\sqrt{m_{\pi}^{ 2}\,
+p^2}.
 \ee
The  pressure of the kaon gas is given as follows
 \be P_{{K}}^{\rm part}&=& P_{K^+}+ P_{K^0}+P_{K^-}+P_{\bar{K}^0}\\
&=&\frac{1}{3}\intop_0^{\infty}\frac{\rmd p \ p^4}{2\pi^2}\,
\left[
 \frac{f_{K^+} (\om_{K^+}(p))}{\om_{K^+}(p)}+\frac{ f_{K^0}
(\om_{K^0}(p))}{\om_{K^0}(p)}\right] \nonumber\\
&+&\frac{1}{3}\intop_0^{\infty}\frac{\rmd p p^4}{2\pi^2}\, \left[
\frac{ f_{K^-} (\om_{K^-}(p))}{\om_{K^-}(p)}+\frac{ f_{\bar{K}^0}
(\om_{\bar{K}^0}(p))}{\om_{\bar{K}^0}(p)}\right] ~, \nonumber
 \ee
where
 \be \label{EK+} \om_{K^{\pm}}(p)= \pm g_{\om K}^{*} \ \om_0
\pm g_{\rho K}^{*} \ R_0 + \sqrt{m_{K}^{* 2}\, +p^2} ~,\quad
m_{K}^{*}=m_{K}-g_{\sigma K}^{*}\sigma
 \ee
and $g_{mK}^{*}=g_{mK}\chi_m$ for the parameters of the set $B$
\cite{KTV},  we use here. Note that in neglecting Coulomb effects
the energy of $K^{+}$ coincides with that for $K^{0}$ and the
energy of $K^{-}$ coincides with  the $\bar{K}^{0}$ energy.

The $\eta$-contribution to the pressure is given by
 \be
P_{\eta}^{\rm part}&=& \frac{1}{3}\intop_0^{\infty}\frac{\rmd p \
p^4}{2\pi^2}\, \frac{ f_{\eta} (\om_{\eta}(p))}{\om_{\eta}(p)}~,
\quad \omega_{\eta}=\sqrt{m_{\eta}^{* 2}+{p}^2}
 \ee
with
 \be \label{etam} m_{\eta}^{* 2}=\left(
m_{\eta}^{2}-\sum_{b\in\{b\}} \frac{\Sigma_{\eta b}}{f^2_{\pi}}
\left<\bar\Psi_b \,\Psi_b\right> \right) / \left(
1+\sum_{b\in\{b\}}\frac{\kappa_{\eta
b}}{f^2_{\pi}}\left<\bar\Psi_b \,\Psi_b\right>\right)
 \ee
expressed in ref. \cite{ZPLN} in terms of
 the total baryon scalar density
$\sum_{b\in\{b\}} \ n_{sb}= \\ \sum_{b\in\{b\}}\left<\bar\Psi_b
\,\Psi_b\right>$.

In the above equations the Bose distributions of
 excitations are
 \be
\label{Tbos} f_{\rm i} &=&\frac{1}{\mbox{exp}[(\sqrt{m_{\rm
i}^{*2}+p^2}- \mu_{\rm i}^*)/T]-1},\\ \label{Tmubos} \mu_{\rm i}^*
&=&\mu_{\rm i} +Q_{\rm i} \mu_{\rm
  ch}-Q^{\rm vec}_{\rm i} \ g_{\om {\rm i}}^{*} \
\om_0 -Q^{\rm vec}_{\rm i} g_{\rho {\rm i}}^{*} R_0 ,\nonumber\\
i\in\{{\rm bos.ex}\}&=& \sigma^{'} ,\om^{'} ,\rho^{' +} ,\rho^{'
0} , \rho^{' -}  ; \pi^{+},\pi^0 ,\pi^- ; K^+ ,K^0 , K^-
,\bar{K}^0 ; \eta ;\nonumber\\
 && K^{*+} ,K^{*0} , K^{*-} ,
\bar{K}^{*0}; \eta^{'}; \varphi ~.\nonumber
 \ee
Here $\mu_i =\mu_{\rm str}$  for strange particles $K$ and $K^{*}$
and $\mu_i =-\mu_{\rm str}$ for their anti-particles; $Q_{\rm i}$
is the boson electric charge in proton charge units (ignoring
Coulomb effects we put $Q_i =0$), $Q^{\rm vec}_{\rm i} =+1$ for
particle, $Q^{\rm vec}_{\rm i} =-1$ for antiparticle and $Q^{\rm
vec}_{\rm i} =0$ for "neutral" particles (with all zero charges
including strangeness).
We take couplings as in \cite{KTV}, $g_{mi}^{*}=0$ for all $i$
except $i=K, \eta$. For $\eta$ we select the parameter choice
$\Sigma_{\eta b}= 140$~MeV, $\kappa_{\eta b}=0.2$. As in
\cite{KTV} we assume $m^{*}_{K^{*}}=m_{K^{*}}$,
$m^{*}_{\eta^{'}}=m_{\eta^{'}}$, $m^{*}_{\varphi}=m_{\varphi}$ due
to the absence of the corresponding experimental data.  In this
paper, we will focus on the consideration of the
isospin-symmetrical matter; therefore, we may put $R_0 =0$.

The density of the gas of Bose excitations of the given species
$i$ is determined by the integral
 \be
\label{bosun} n_i =g_i \intop_0^{\infty}\frac{\rmd p \
p^2}{2\pi^2}\, f_i(p)~, \quad i\in\{{\rm bos.ex}\},
 \ee
where $g_i$ is the corresponding degeneracy factor.

Note that  the $p$-wave pion and kaon terms can easily  be
included. For that one  needs to replace $\om_{\pi}(p)$ and
$\om_{K}(p)$ with more complicated expressions, see refs.
\cite{MSTV90,KV03}. In order to obtain the
temperature-density dependent pion and kaon spectral functions one
needs to calculate  the pion and the kaon self-energies including
baryon-baryon hole loops and baryon-baryon correlation effects.
These effects may result in appearance of the pion and antikaon
condensates in dense and not too hot nuclear matter. Within the
SHMC model we incorporate only interactions of particles through
mean fields. Even in this case many coupling constants are not
well fixed due to the lack of experimental data. Thus, we postpone
the inclusion of the $p$-wave interactions to the future work.

{\em{A more nontrivial task is to fix  the terms $P_{\sigma}^{\rm
part}$, $P_{\om}^{\rm part}$, $P_{\rho}^{\rm part}$.}} For the
$\sigma^{'}$, $\om^{'}$, $\rho^{'}$ contributions to the pressure
we use the standard ideal gas expressions with effective masses
determined as follows. Let us first focus on the contribution of
the $\sigma^{'}$-excitations. In order to get $P_{\sigma}^{\rm
part}$, we should expand total pressure $P [\sigma ,\omega_0
(\sigma ) ,R_0 (\sigma )]$  in $\sigma^{'} =\sigma -\sigma^{\rm
cl}$. The term linear in $ \sigma^{'}$ does not give a
contribution due to subsequent requirement of the pressure minimum
in $\sigma^{\rm cl}$. The quadratic term produces
 effective $\sigma$ particle mass squared,
 \be\label{Pspa1}
 (m_{\sigma}^{\rm part*})^2 =-\frac{{d^2 P [\sigma ,\omega_0 (\sigma ),R_0
(\sigma)]}}{{d \sigma^2}} =-\frac{{d^2 P [f ,\omega_0 (f ),R_0
(f)]}}{{d f^2}}\left(\frac{d f}{d \sigma}\right)^2 .
 \ee
The first-order derivative  $dP/df=0$, as it follows from the
equations of motion. Since we deal with the strong interaction
problem and the general solution is impossible, one should use
some approximations. First, keeping only quadratic terms in all
thermodynamical quantities in boson fluctuating fields we
disregard boson excitation contributions to the $\sigma^{'}$,
$\om^{'}$, $\rho^{'}$ effective masses (higher order effects in
boson fluctuating fields). Within this approximation we neglect
the $P_{\rm bos.ex.}$ term in (\ref{Pspa1}). Moreover, our aim is
the construction of a thermodynamically consistent model where the
baryon density and  the entropy density, calculated by using the
thermodynamic relation (\ref{cons1}) and relation
 \be\label{consist} S=\left.\frac{\partial P}{\partial
T}\right|_{\mu}~,
 \ee
coincide with  quasiparticle expressions   (\ref{bari}) and
 \be\label{entrQP}
  &&S=S_{\rm bar} +S_{\rm bos.ex.}, \quad
S_{\rm bar} =\sum_{b\in\{b\}} g_b\int \frac{\rmd p_b \
p_b^2}{2\pi^2}[-(1-f_b )\mbox{ln} (1-f_b )-f_b \mbox{ln} f_b
],\nonumber\\ &&S_{\rm bos.ex.}=\sum_{i\in\{\rm bos.ex.\}}g_i \int
\frac{\rmd p_i \ p_i^2}{2\pi^2}[(1+f_i )\mbox{ln} (1+f_i )-f_i
\mbox{ln} f_i ],
 \ee
namely, the latter expressions  in ref. \cite{KTV}. In order to
keep thermodynamic consistency of the model, we suppress the
baryon contribution in (\ref{Pspa1}). This term is associated with
taking into account of the  baryon-antibaryon and baryon-baryon
hole loops. Problems arisen, if one includes baryon loops, are
discussed in Appendix B.

Thus, the squared effective mass of the  $\sigma^{'}$ excitation
will be given by the expression
 \be \label{spa3} (m_{\sigma}^{\rm part*})^2
\simeq - \sum_{m\in\{m\}}\frac{{d^2 P_m^{\rm MF}[f ,\omega_0 (f
),R_0 (f)]}}{{d f^2}}\left(\frac{d f}{d \sigma}\right)^2 .
 \ee
 Using the relation $\chi_{\sigma}=\Phi
=1-f$ we find that
 \be\label{h}\frac{d f}{d
\sigma}=\frac{g_{\sigma N}}{m_N}(1-f)^2 .
 \ee
For the isospin-symmetrical nuclear matter $N=Z$,
we obtain
 \be
\label{spa4} (m_{\sigma}^{\rm part*})^2 =\left[
U^{''}_{f}+\frac{m_N^4}{C_{\sigma}^2}\right] \left(\frac{g_{\sigma
N}}{m_N}\right)^2(1-f)^4 .
 \ee
One could use a different approximation introducing the effective
mass term with the help of the expression for the Hamiltonian
(Lagrangian). In the latter case, baryon terms are added to the
meson mean field contribution and derivatives of the Hamiltonian
are taken at fixed $\Psi_B$. In this case
 \be
\label{spa2} (m_{\sigma}^{\rm part*})^2 =\left<\frac{{d^2 H}}{{d
\sigma^2}}\right>=  \left<\frac{{d^2 H}}{{d f^2}}\right>
\left(\frac{d f}{d \sigma}\right)^2 ,
 \ee
 where $H$ is the  Hamiltonian and the averaging procedure
is carried out over a thermal equilibrium state after taking the
derivatives. We used that $\left<\frac{{d H}}{{d \sigma}}\right>=
0$, as it follows from the equations of motion. Neglecting the
meson excitation terms in $H$ we find
 \be \left<\frac{{d^2 H}}{{d f^2}}\right>&=&-
\sum_{m\in\{m\}}\frac{{d^2 P_m^{\rm MF}[f ,\omega_0 (f ), T]}}{{d
f^2}}+ \left<\sum_{b\in\{b\}}\frac{{d^2 H_b [f ,\omega_0 (f ),
T]}}{{d f^2}}\right>\nonumber\\ &=&
U^{''}_{f}+\frac{m_N^4}{C_{\sigma}^2}+\frac{C_{\om}^2}
{2m_N^2}\left(\frac{d^2}{df^2}\frac{1}{\eta_{\om}}\right)\sum_b
x_{\om b} \ n_{sb},
 \ee
where the baryon (antibaryon)  scalar density $n_{sb}$ is given by
 \be
n_{sb}= (2s_b +1) \int_{0}^{\infty}\frac{dp
p^2}{2\pi^2}\frac{m_b^{*}}{\om_b}f_b . \nonumber
 \ee
Since in ref. \cite{KTV} the value $\eta_{\om}^{-1}$ was chosen as
a linear function of $f$,
 \be
\eta_{\om}^{-1}=(1+zf)/(1+zf(n_0)), \quad z=const,
  \ee
($z=0.65$ in the KVOR-based SHMC model), we get
$\sum_{b\in\{b\}}\frac{{d^2 H_b [f ,\omega_0 (f ), T]}}{{d
f^2}}=0$, and both results (\ref{spa3}) and (\ref{spa2}) coincide.
From here it is seen that in our approximation $(m_{\sigma}^{\rm
part*})^2$ does not explicitly depend on the baryon variables and
 the temperature. Obviously, {\em{ within this approximation our
model remains thermodynamically consistent}} since meson
excitation terms in the pressure do not contribute to the
derivatives of the pressure over the baryon chemical potential (to
obtain $n_B$) and over the temperature (to get entropy $S$).

Note that in \cite{KTV} we introduced the squared effective mass
of the $\sigma^{'}$-excitation as the second derivative of the
energy density. Since at the same time  the contributions of all
baryon particle-particle hole and particle-antiparticle (loop)
terms were suppressed, this result  coincides with expressions
(\ref{spa3}) and (\ref{spa2}).

Within our approximation the masses of vector particles are given
by
 \be \label{spaom} &&(m_{\om_i}^{\rm part*})^2
=\left<\frac{{\partial^2 H}}{{\partial \om_i^2}}\right>=-
\sum_{m\in\{m\}}\frac{{\partial^2 P_m^{\rm MF}[f ,\omega_0
]}}{{\partial \om_i^2}},\\ &&(m_{\rho_i}^{\rm part*})^2
=\left<\frac{{\partial^2 H}}{{\partial R_i^2}}\right> =-
\sum_{m\in\{m\}}\frac{{\partial^2 P_m^{\rm MF}[f ,\omega_0
]}}{{\partial R_i^2}}, \quad i=0,1,2,3.\nonumber
 \ee
 In this case zero-components of one $\om$
and three $\rho$-excitations prove to be the same as those
following from the mean-field mass terms
 \be
\label{omr} m_\om^{\rm part
  *} =m_\om |\Phi_\om (f)|, \quad m_\rho^{\rm part
  *} =m_\rho |\Phi_\om (f)|~.
 \ee
Moreover, there are excitations of two magnetic-like components of
$\om$ and six magnetic-like components of $\rho$. Their masses are
given by eqs. (\ref{omr}) since there are no $\vec{\om}$ and
$\vec{R}$ mean fields, namely, expressions (\ref{omr}) were used
in \cite{KTV}.

{\em{Thus, neglecting the baryon  and boson excitation
contributions to the effective meson excitation masses we obtain a
rather simple thermodynamically consistent description.}}

\subsection{Some numerical results}

{In Figs. \ref{KTV}, \ref{pT4rcmp} we demonstrate the size of
corrections which arise provided  boson excitations are
incorporated in the equations of motion. Particle excitation
masses are calculated following eq. (\ref{spa3}) for $\sigma$
meson excitation and (\ref{omr}) for the nucleon, $\om$, and
$\rho$ excitations. Figure \ref{KTV} demonstrates the ratio of the
effective masses to the bare masses for the
nucleon-$\om^{'}$-$\rho^{'}$ (left panel) and for $\sigma^{'}$
(right panel) as a function of the temperature at $n_B=0$ (thin
curves) and at $n_B =5~n_0$ (bold lines). Solid curves are
calculations of the given paper, when boson excitations contribute
to the equations of motion, whereas dashed curves show
perturbative calculations of ref. \cite{KTV}. Although for $T\gsim
120$~MeV there appear pronounced differences between both
treatments of excitations, the qualitative behavior remains
unchanged. Effective masses of all excitations exhibit  a similar
behavior as functions of the temperature and the density, in a
line with the mass-scaling hypothesis that we have exploited for
the mean fields.

 Let us also compare our result for $N$, $\om$, $\rho$
effective masses (Fig. \ref{KTV} left) with that previously
obtained in the model of ref. \cite{LKB}, their Fig. 1 left (for
$N$) and right (for $\om, \rho$). Shapes of the curves look
similar. However in our case $N$, $\om$, $\rho$ are scaled by one
scaling function, whereas  ref. \cite{LKB} used different scalings
for $N$ and for $\om, \rho$. In their case the masses drop much
stronger with the density increase but remain finite at high
temperatures ($T\gsim 200$~MeV), whereas in our case they drop to
zero at $T_c \sim 210$~MeV.

\begin{figure}[t]
 \hspace*{16mm}
\includegraphics[width=100mm,clip]{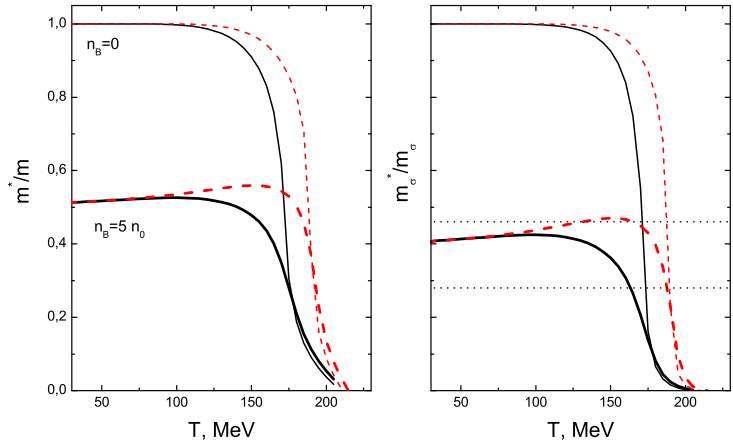}
\caption{ Temperature dependence of the effective-to-bare mass
ratio for nucleon-$\om^{'}$-$\rho^{'}$ (left panel) and for
$\sigma^{'}$ (right panel) for $n_B=0$ and $n_B=5~n_0$. The solid
lines are our results, the dashed ones are perturbative
calculations  from \cite{KTV}. To guide the eye, the horizontal
dots show the $m_\sigma^* =2m_{\pi}$ and $m_\sigma^* =m_{\pi}$
thresholds.}
 \label{KTV}
\end{figure}
\begin{figure}[thb]
 \hspace*{20mm}
\includegraphics[width=80mm,clip]{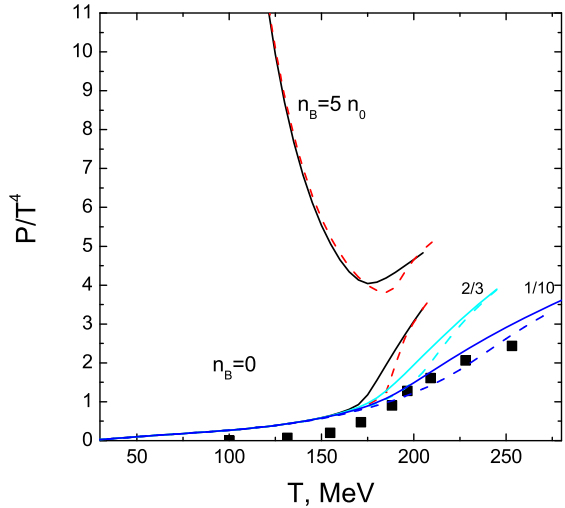}
\caption{ Temperature dependence of the reduced pressure. Solid
lines show calculations of the given work, whereas dashed lines
demonstrate the results of the perturbative treatment of boson
excitations \cite{KTV}. The curves labeled by $2/3$ and $1/10$
correspond to the case when all $g_{\sigma b}$ couplings except
for nucleons are suppressed by factors $2/3$ and $1/10$,
respectively. Filled squares show the lattice QCD result for the
2+1 flavor case \cite{Karsh}.  }
 \label{pT4rcmp}
\end{figure}

The effect of the mentioned nonperturbative treatment of boson
excitations on thermodynamic quantities is presented in Fig.
\ref{pT4rcmp}. The temperature dependence is shown for the reduced
pressure calculated in the given work (solid lines) and in the
perturbative treatment of boson excitations \cite{KTV} at $n_B =0$
and $n_B =5n_0$. One can see that the differences are rather
noticeable only in the temperature interval $170<T<200$~MeV,
provided the baryon-meson couplings are not artificially
suppressed. In the case ($n_B =0$), when all $g_{\sigma b}$
couplings except for nucleons are artificially suppressed by
factors $2/3$ and $1/10$, \footnote{As in Fig. 10 of  \cite{KTV},
we suppress $g_{\sigma b}$-couplings, not $g_{m b}$, as mistakenly
indicated there.} the temperature interval, where solid and dash
curves deviate from each other, is broader but the value of the
deviation is smaller. The curve for $g_{\sigma b}$ suppressed by
$1/10$ reasonably matches the lattice data for $T_c < T\lsim
230$~MeV. As has been mentioned in \cite{KTV}  one could fit
the lattice data even in a broader region of temperatures (up to
$500~$MeV) if one introduced $\chi_\sigma <\Phi$.
 A violation of the universality of the $\sigma$ scaling
would be also  in a line with that we have used for $\om$ and
$\rho$, $\eta_{\om}\neq 1$ and $\eta_{\rho}\neq 1$, as required to
describe properties of neutron stars, cf. \cite{KV03}. However we
will not elaborate such a possibility in the present work. Instead
we simply suppress $g_{\sigma b}, b\neq N$. Hydrodynamic
simulations of heavy ion collisions are fully determined by the
EoS to be described with quark and gluon degrees of freedom or
with only hadron ones. With such an  EoS (with suppressed
$g_{\sigma b}$) a quark liquid would masquerade as a hadron one.

\begin{figure}[thb]
 \hspace*{2mm}
\includegraphics[width=130mm,clip]{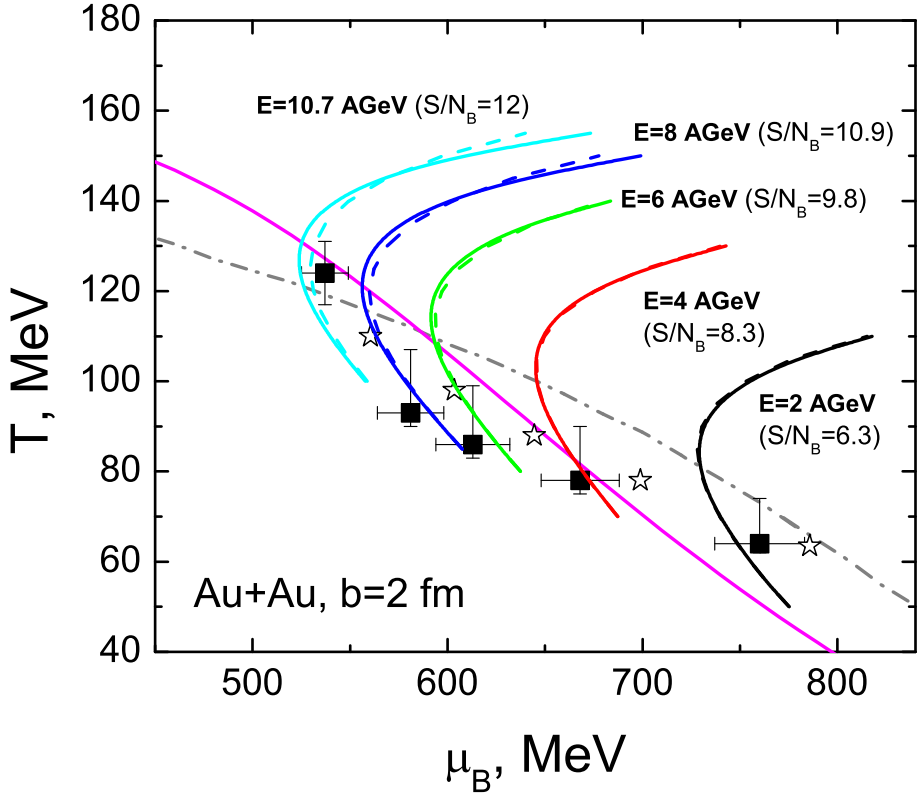}
\caption{ Isentropic trajectories for central Au+Au collisions at
different bombarding energies calculated in the present paper
(solid lines) and with the perturbative treatment  of boson
excitations \cite{KTV} (dashed lines). The results  are presented
for the SIS-to-AGS energies. Experimental points with error bars
are taken from~\cite{ABMS05}. The freeze-out points marked by
stars are obtained in \cite{KTV}. Thin line corresponds to the
freeze-out curve in~\cite{ABMS05} while dash-dotted line is that
from~\cite{CR98}. }
 \label{Trad}
\end{figure}

\begin{figure}[thb]
 \hspace*{2mm}
\includegraphics[width=130mm,clip]{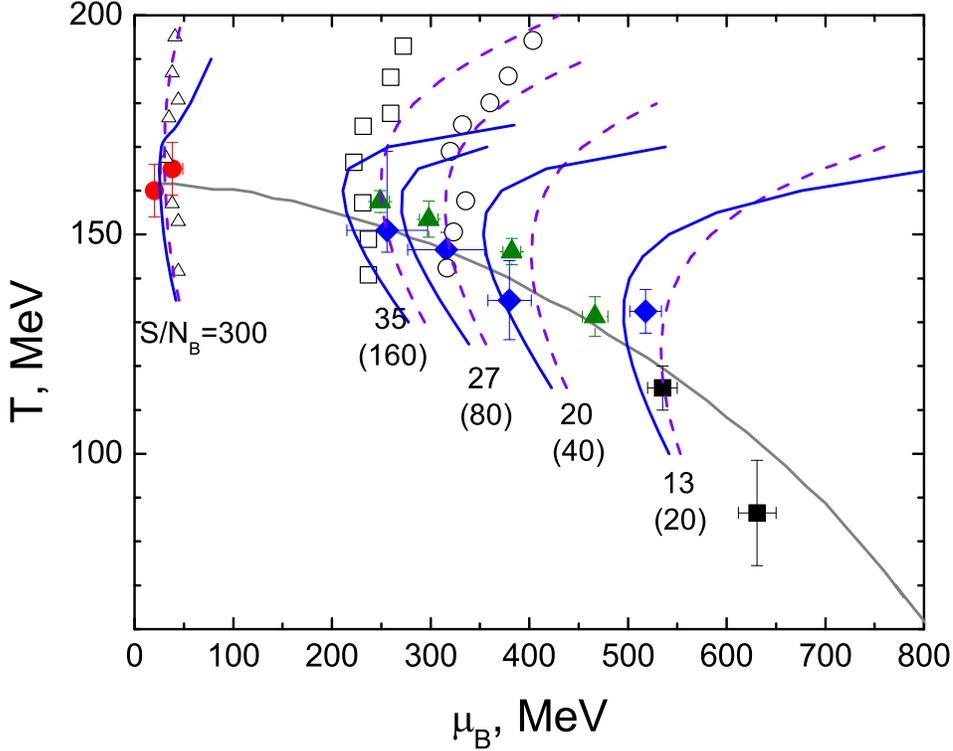}
\caption{Isentropic trajectories for central Au+Au collisions at
different bombarding energies calculated in the present paper.
Calculations are performed from AGS  to RHIC energies for not
suppressed couplings (solid lines) and with suppressed $g_{\sigma
b}$  couplings (except for nucleons) by a factor of $1/10$
(dashed lines). The filled diamonds and triangles are obtained
from the $4\pi$ particle ratios in~\cite{ABMS05} and \cite{BMG06},
respectively. Filled circles are RHIC data based on the middle
rapidity particle ratios~\cite{ABMS05}. Open circles, squares and
triangles are the lattice 2-flavor QCD results~\cite{EKLS05} for
$S/N_B=$ 30, 45 and 300, respectively. The thin  line corresponds
to the freeze-out curve in~\cite{ABMS05}.}
 \label{Trad1}
\end{figure}

In Fig. \ref{Trad} we show the isentropic trajectories for central
Au+Au collisions at different bombarding energies calculated in
the present paper (solid lines) and with the perturbative
treatment  of boson excitations from ref. \cite{KTV} (dashed
lines).  As we can see from Fig. \ref{Trad}, for AGS energies
the trajectories calculated here almost coincide with those
computed in ref. \cite{KTV} where boson excitations were treated
perturbatively.  Then, as in \cite{KTV} we extend our analysis to
higher bombarding energies. The entropy per baryon participants
was calculated in~\cite{RDB98} within the 3-fluid hydrodynamic
model assuming occurrence of the first order phase transition to a
quark-gluon plasma. The energy range from AGS to SPS was covered
there. We use the $S/N_B$ values for $E_{lab}=$158, 80, 40 and 20
AGeV, at which the particle ratios were measured by the NA49
collaboration (cf.~\cite{ABMS05}). At the RHIC we put $S/N_B=300$
in accordance with an estimate in~\cite{EKLS05}.

Since  there is a reasonable fit of the lattice data on the
pressure for $T\gsim 200$~MeV with reduced meson-baryon coupling
constants, see \cite{KTV} and  Fig. \ref{pT4rcmp}, for high
collision energies  we show in Fig. \ref{Trad1} both the results,
when couplings are not suppressed (solid curves) and also when all
$g_{\sigma b}$ couplings except for nucleons are suppressed by a
factor of $1/10$. As is seen, the decrease in $g_{\sigma b}$
improves the high-temperature description (as compared to
the case of not suppressed couplings) for the lattice data at high
temperatures not only for the case $\mu_b =0$ but also for $\mu_b
\neq 0$.  For  low temperatures the solid and dashed curves
(the curves for not suppressed and suppressed $g_{\sigma b}$
couplings, respectively) are very close to each other.  Also note
that one should be rather critical performing comparison with the
existing lattice data, especially for the case $\mu_b \neq 0$
(presented in Fig. \ref{Trad1} for 2 flavors). Doing this
comparison we tentatively hope that results of future more
realistic lattice calculations will not deviate much from the
existing ones.

 In all cases our new calculations presented in  Fig.
\ref{Trad1} (if one additionally suppresses couplings), as well as
corresponding perturbative calculations \cite{KTV}, describe
reasonably the lattice data on the $T-\mu_B$ trajectories and
freeze out points.

Concluding, our improvement of the model with keeping boson
excitation terms in the equations of motion neither spoils nor
improves the agreement with the lattice results and with the
thermodynamic parameters extracted at freeze out. Thus, both
perturbative and nonperturbative treatments of boson excitation
effects  can be used with equal success. However, one should note
that thermodynamic consistency conditions are fulfilled exactly in
the non-perturbative treatment of the given work, whereas in the
perturbative treatment of \cite{KTV} they were satisfied only
approximately.

\section{Inclusion of finite resonance widths}\label{res}

In the framework of our model we treated all resonances as
quasiparticles neglecting their widths. Excitations interact
only with the mean fields. Thus imaginary parts of the
self-energies as well as some contributions to the real parts  are
not taken into account. Definitely it is only a rough
approximation carried out just for simplification.

 At the resonance peak   the vacuum $\Delta$-isobar mass width
is $\Gamma_{\Delta}^{\rm max}\simeq 115$~MeV. In reality
$\Gamma_{\Delta}$ is the temperature-, density- and
energy-momentum-dependent quantity. For low $\Delta$-energies the
width is much less than $\Gamma_{\Delta}^{\rm max}$. A typical
$\Delta$ energy is $\om_\Delta-m_\Delta^*\sim T$. Thus for low
temperature, $T\lsim \epsilon_F$ ($\epsilon_F$ is the nucleon
Fermi energy) the effective value of the $\Delta$-width is
significantly less than $\Gamma_{\Delta}^{\rm max}$. At these
temperatures the quasiparticle approximation does not work for
$\Delta$'s but their contribution to thermodynamic quantities is
small. When the temperature is  $\gsim m_\pi$  there appears
essential temperature contribution to the width and the resonance
becomes broader \cite{VS91}. At such temperatures $\Delta$'s
essentially contribute to thermodynamic quantities. Only for
temperatures $T\gsim \Gamma_{\Delta}^{\rm max}(T)$ the
quasiparticle approximation becomes a reasonable approximation.

The $\rho$- and $\sigma$-mesons also have rather broad widths.
E.g., at a maximum  the $\rho$-meson width is about $150$~MeV. The
observed enhancement of the dilepton production at CERN, in
particular in the recent NA60 experiment~\cite{NA60} on
$\mu^+\mu^-$ production, can be explained by significant
broadening of the $\rho$ in matter \cite{RW}, though decreasing of
the $\rho$ mass could also help in explanation of the data
\cite{TS}\footnote{As demonstrated in \cite{TS}, the
calculated large mass shift is mainly caused by the assumed
temperature dependence of the in-medium mass.
Inclusion of this temperature dependence
modifies the scaling hypothesis originally claimed by Brown and
Rho. Some arguments on what the proper mass-scaling  predicts for
dilepton production in HIC, e.g. NA60, were given in
\cite{BR06}.}. Besides, as was shown in \cite{RZM} to be
consistent with the QCD sum rules, both the collisional broadening
and dropping of the $\rho$ mass should be taken into account. Even
more, one should be careful with interpretation of the NA60
experiment: Dileptons carry direct information on the $\rho$ meson
spectral function only if the vector dominance is valid but
generally it is not the case \cite{Bea08}.

Also particles which have no widths in vacuum like nucleons
acquire the widths in matter due to collisional broadening. Their
widths grow with   the temperature increase, cf. \cite{V04}. As we
have argued above in case with $\Delta$ isobars, the quasiparticle
approximation may become a reasonable approximation at
sufficiently high temperature, if $T\gsim \Gamma (T)$.

Let us  roughly estimate the effect of finite baryon and meson
resonance widths on particle distributions and the EoS in order to
understand how much and in which temperature-density regions these
effects may affect  results calculated within our quasiparticle
SHMC model.

It is convenient to introduce single-particle spectral $
\widehat{A}_{\rm r}$ and width $\widehat{\Gamma}_{\rm r}$
functions (operators), cf. \cite{V04},
 \be
\label{A-G} \widehat{A}_{\rm r} =-2\im \widehat{G}^R_{\rm r}
(q)=-2\im\frac{1} { \widehat{M}_{\rm r} +i\widehat{\Gamma}_{\rm r}
/2},\,\,\, \ \  \widehat{\Gamma}_{\rm r} =-2\im
\widehat{\Sigma}^R_{\rm r} \,,
 \ee
 where $\widehat{G}_{\rm
r}^R (q)$ is the full retarded  Green function of the fermion and
$\widehat{\Sigma}^R_{\rm r}$ is the retarded self-energy. The
quantity
 \be\label{M} \widehat{M}_{\rm r}
=(\widehat{G}^{0,R})^{-1}-\re \widehat{\Sigma}^R_{\rm r}
 \ee
demonstrates  the deviation from the mass shell: $\widehat{M}_{\rm
r}=0$ on the quasiparticle mass shell in the matter,
$\widehat{G}^{0,R}$ is the free Green function. Following
definition (\ref{A-G}) the spectral function of the fermion has
dimensionality of $m^{-1}$ and the width, dimensionality $m$,
whereas for bosons the spectral function  has dimensionality of
$m^{-2}$ and  the width, dimensionality $m^2$.

Let us start with the consideration of a fermion spin $1/2$
resonance  (superscript $\rm f$). The spectral function
satisfies the sum rule:
 \be\label{fsum-r}
\frac{1}{4} \mbox{Tr} \int^{\infty}_0 \gamma_0
\left[\widehat{A}^{\rm f}_{\rm r (+)} (q_0 ,\vec{q})+
\widehat{A}^{\rm f}_{\rm r (-)} (q_0 ,-\vec{q})\right] \frac{d
  q_0}{2\pi}=2,\,\,\,
 \ee
 $\gamma_0$ is the corresponding Dirac matrix, and subscripts $(\pm
)$ specify particle and antiparticle terms.
 The trace is taken over spin degrees of freedom.

Simplifying the spin structure, as it is seen from (\ref{A-G}), we
can present

 \be\label{an} \frac{1}{4} \mbox{Tr}\gamma_0 \widehat{A}^{\rm f}_{\rm r
(+)} =A^{\rm f}_{\rm r}=\widetilde{A}_{\rm r}\om .
 \ee
Dealing further with $A^{\rm f}_{\rm r}=\widetilde{A}_{\rm r} \
\om$ we may not care anymore about the spin structure.  The
$\Delta$ spin $3/2$ resonance can be considered in a similar way,
see \cite{Leupold}.

It was   argued in refs. \cite{IKV1} that at a self-consistent
description the particle densities are given by the Noether
quantities:
 \be\label{NoethB}
 n^{\rm f}_{\rm r}=N_r \int_{0}^{\infty}\frac{4\pi p^2 d
p}{(2\pi)^3}\int_{0}^{\infty}\frac{ d \om}{2\pi} A^{\rm f}_{\rm r}
f^{\rm f}_{\rm r}, \quad f^{\rm f}_{\rm r}=\frac{1}{e^{(\om
-\mu_{\rm r}^{*})/T}+1}.
 \ee
Here $\mu_{\rm r }^{*}=\mu_{N}^{*}$ for a nucleon resonance such
as $\Delta$.  The antiparticle density is obtained from
(\ref{NoethB}) with the help of the replacement $\mu_{\rm
r}^{*}\rightarrow -\mu_{\rm r}^{*}$. The baryon density of the
given species is  then  $n_{\rm r}^{\rm bar} =n_{\rm r}(\mu_{\rm
r}^{*})-n_{\rm r}(-\mu_{\rm r}^{*})$. $N_{\rm r}$ is the
degeneracy factor.

To do the problem  tractable, instead of solving a complete set of
the Dyson equations, we may select a simplified phenomenological
expression for $A^{\rm f}_{\rm r}$ (compare with \cite{Leupold}),
e.g.,
 \be\label{fenomen} A^{\rm f}_{\rm r} =\frac{2\xi \om [2\widetilde{\Gamma}_{\rm r}
(s)+2\delta] }{(s -m^{*2}_{\rm r})^2 +[\widetilde{\Gamma}_{\rm r}
(s)+\delta]^2 }, \quad \xi =const,\quad s=\om^2 -p^2
>0,
 \ee
with  $\delta \rightarrow +0$.  The value $\delta \rightarrow
0$ is introduced to easier extract the quasiparticle term.
 Separating out the quasiparticle
pole we obtain
 \be
 \label{fenomen1} A^{\rm f}_{\rm r} =\frac{2\xi \om [2\widetilde{\Gamma}_{\rm r}
(s)] }{(s -m^{*2}_{\rm r})^2 +[\widetilde{\Gamma}_{\rm r} (s)]^2
}+2\xi \om\cdot 2\pi\delta (s-(m_{\rm r}^*)^2 ) \ \theta (s_{\rm
th}-(m_{\rm r}^*)^2 ).
 \ee
 Here $s_{\rm th}$ is the resonance threshold value of $s$. Note
that within our ansatz, the spectral function depends only on the
$s$-variable. It might be the case only for a dilute matter, when
the density and temperature dependence of the width is rather
weak. Furthermore, instead of a calculation of the
density-temperature dependent part of the width, which actually
can't be performed in the framework of our model, we vary the
energy dependence and the amplitude of the width thus simulating
in such a way collision broadening effects.

For the decay of the resonance into two particles ($r\rightarrow
1+2$) one may use a simple $s$-variable dependence of the width:
 \be
 \label{widthf}&&\widetilde{\Gamma}_{\rm r} (s)=\Gamma_0 \ m_{\rm r} \ F(s)
\left(\frac{p_{\rm c.m.}^2 (s, m_1^* , m_2^* )}{p_{\rm c.m.}^2
(m_{\rm r}^2 , m_1 , m_2 )}\right)^{\alpha}\theta \left(s-(m_1^*
+m_2^* )^2\right),\\ &&p_{\rm c.m.}^2 (s, m_1^* , m_2^* )=
\frac{(s-(m_1^* +m_2^*)^2 )(s-(m_1^* -m_2^* )^2 )}{4s}
.\nonumber
 \ee
 Here $\Gamma_0 =const$,  the width tends to zero at
the threshold $s\rightarrow s_{\rm th}=(m_1^* +m_2^*)^2$, $\alpha
=l+1/2$, with $\alpha =1/2$ for $s$ and $\alpha$ =3/2 for $p$
resonance. An extra form-factor, $F(s)$, is introduced to correct
the high-energy behavior of the width.

To simplify expression (\ref{widthf}), we may expand
$\widetilde{\Gamma}_{\rm r} (s)$ near the threshold transporting
remaining $s$-dependence to the form factor:
 \be
 \label{widthf1}\widetilde{\Gamma}_{\rm r} (s)=\Gamma_0 \ F(s) \ m_{\rm r}
\left(\frac{s^{1/2}-s_{\rm th}^{1/2}}{m_{\rm r}-s_{\rm th}^{1/2}}
\right)^{\alpha}\theta (s-s_{\rm th})~,
 \ee
where one can take
 \be \label{FF}
F=\frac{1}{1+[(s-s_{th})/s_0]^\beta}
 \ee
 with $s_0$ and $\beta$ being
constants. The parameters can be fitted to satisfy  experimental
data.

The energy dependence of the width causes a problem. With a simple
ansatz for the behavior $\widetilde{\Gamma}_{\rm r} (s)$ we get a
complicated $m^{*2}_{\rm r} (s)$ dependence, as it follows from
the Kramers-Kronig relation. However, since  $m^{*2}_{\rm r} (s)$
is a smooth function of $s$, one may ignore this complexity taking
for simplicity $m^{*2}_{\rm r}$ as a constant. The factor $\xi$ is
introduced to fulfill the sum-rule:
 \be \int_{0}^{\infty}\frac{
\rmd s }{4\pi} \widetilde{A}_{\rm r} =1 ,
 \ee
that yields $\xi \simeq 1+O(\Gamma_0 /m^{*}_{\rm r} )$ (for
$m^{*}_{\rm r}\gg \Gamma_0 $), $m_{\rm r}^* >m_1^* +m_2^*$. In the
case of $m_{\rm r}^* <m_1^* +m_2^*$ there appears an extra
quasiparticle term in the spectral function, that contributes to
the sum-rule.
\begin{figure}[thb]
 \hspace*{2mm}
\includegraphics[width=130mm,clip]{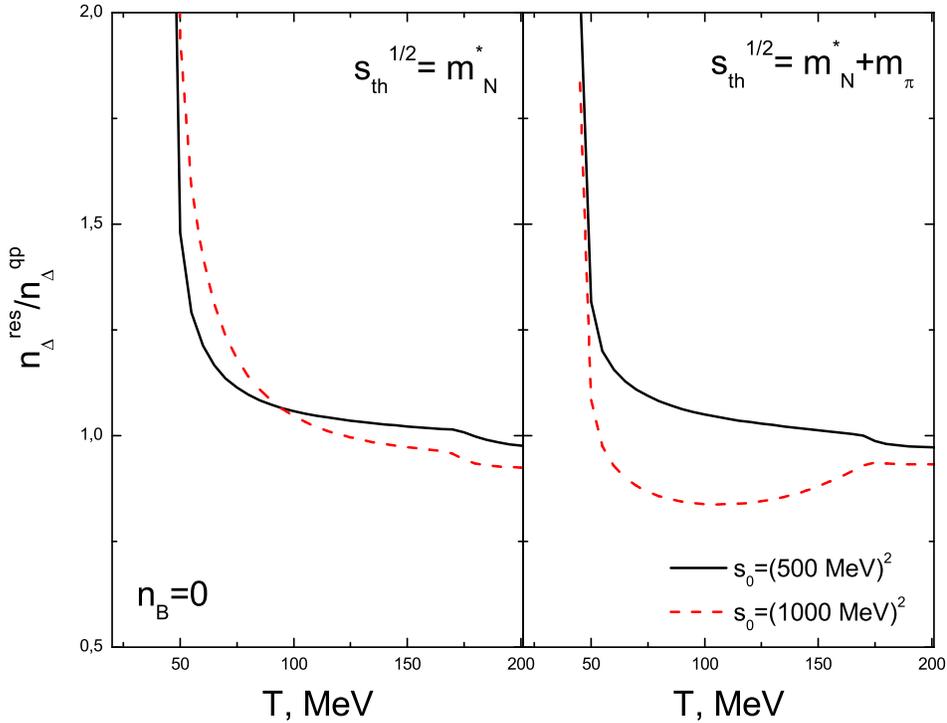}
\caption{ Ratio  of the $\Delta$ isobar density, calculated with
the inclusion of the width, to that of the quasiparticle one,
$R_{\Delta}=n^{\rm res}_{\Delta}/n_{\Delta}^{\rm qp}$, as a
function of temperature at $n_B =0$ for two values of the
resonance threshold energy $s_{th}^{1/2}$. The parameters of
calculation are presented in the figure.
 }
 \label{ratiosD1}
\end{figure}

\begin{figure}[thb]
 \hspace*{2mm}
\includegraphics[width=130mm,clip]{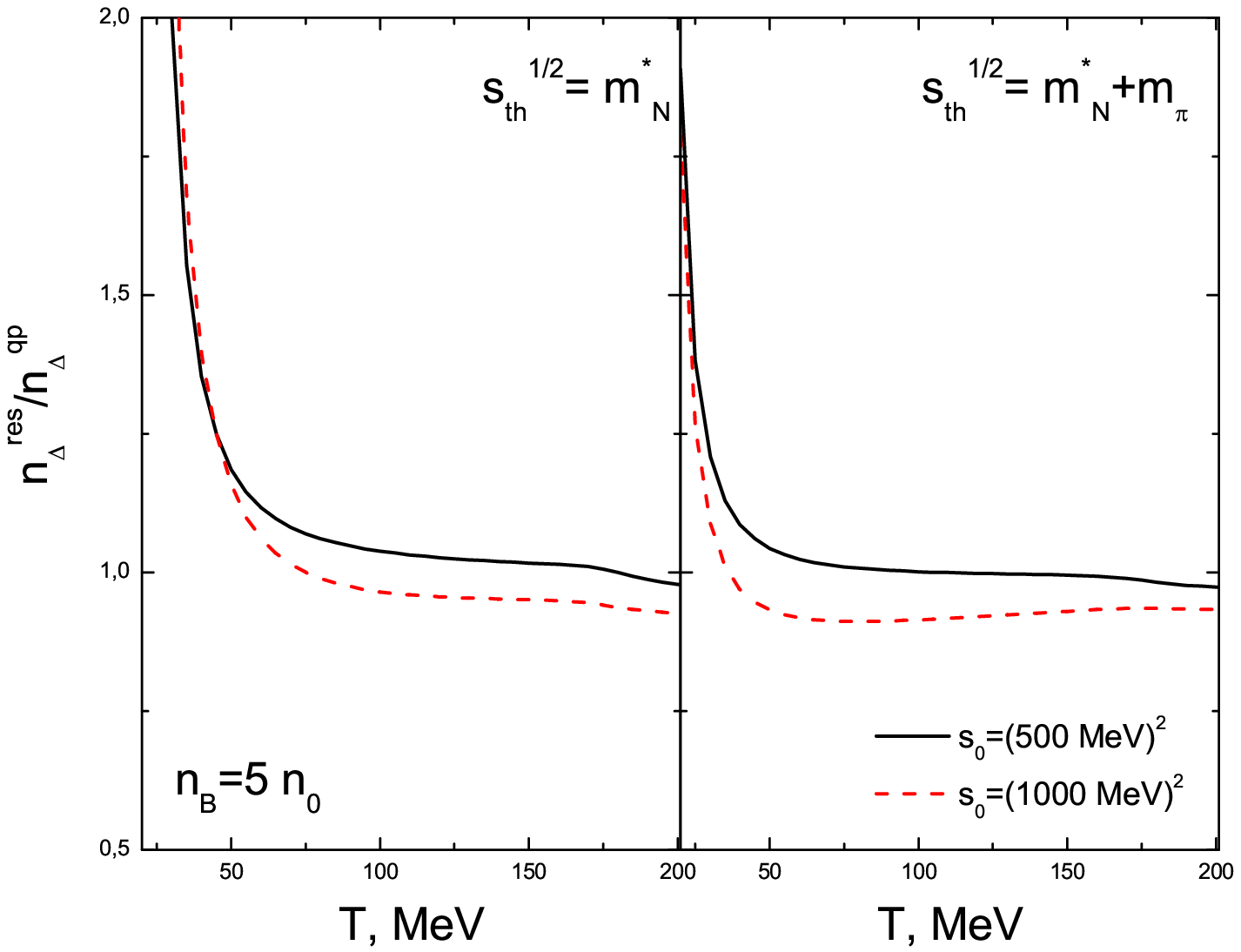}
\caption{ The same as in Fig. \ref{ratiosD1} but for $n_B =5~n_0$.
 }
 \label{ratiosD11}
\end{figure}

In Figs. \ref{ratiosD1} and \ref{ratiosD11}, we present the ratio
$R_{\Delta}=n^{\rm res}_{\Delta}/n_{\Delta}^{\rm qp}$ of the
$\Delta$-isobar density,   calculated following eq.
(\ref{NoethB}), to the quasiparticle density, as a function of the
temperature. The results are presented for the baryon density $n_B
=0$ (in Fig. \ref{ratiosD1}) and for $n_B =5n_0$ (in Fig.
\ref{ratiosD11}). Our aim here is to demonstrate the effect of a
finite particle width. Therefore, instead of searching for the
best fit of the spectral function to available experimental data
we vary the parameters to show how strongly the density ratio  may
depend on them. We take into account the $p$-wave nature of the
resonance and use $\Gamma_0 =115~$MeV. The threshold quantities
are chosen to be $s^{1/2}_{th}=m_N^*$ (left panels) and
$s^{1/2}_{th}=m_N^* +m_{\pi}$ (right panels). In vacuum $s_{\rm
th}^{1/2}=m_N +m_\pi$. Thus, taking $s_{\rm th}^{1/2}=m^*_N
+m_\pi$ we simulate the vacuum resonance placed in the mean field
(in our model $m_{\pi}^* =m_{\pi}$). With $s_{\rm th}^{1/2}=m^*_N$
we simulate the effect of in-medium off-shell pions (virtual pions
can be produced in matter at any energy). The form factor $F$ is
computed with $\beta =3$ and $s_0=(500~\mbox{MeV})^{2}$ (solid
lines) and $s_0=(1000~\mbox{MeV})^{2}$ (dash lines) to present the
dependence of $R_{\Delta}$ on the high-energy behavior of the
width, which is not well defined even in vacuum.

As we can see, in all examples  the curves are rather  flat in the
temperature range $T\gsim 50$~MeV$\div 100$~MeV. For $n_B =0$, at
$T\simeq 170~$MeV there appears a slight bend associated with a sharp
decrease in the nucleon effective mass for $T\gsim 170~$MeV. For
$n_B =5~n_0$ the bend is smeared. A substantial deviation of the
$R_{\Delta}$ ratio from unity for $T\gsim 50$~MeV$\div 100$~MeV in
the case with the cut-off $s_0=(1000~\mbox{MeV})^{2}$ is due to a
larger width at high resonance energies (for large $s$) compared
to the case with the cut-off $s_0=(500~\mbox{MeV})^{2}$. In the
latter example, the quasiparticle approximation becomes
appropriate for $T\gsim 100$~MeV. On the contrary, for
$s_0=(1000~\mbox{MeV})^{2}$ and $n_B =0$ the quasiparticle
approximation does not work at all. For low temperatures, the
$R_{\Delta}$ ratio is significantly higher than unity and for
higher temperatures it becomes essentially smaller than unity. The
broader the width distribution  is, the smaller $R_{\Delta}$ at
large temperatures. The ratio dependence  on the threshold value
$s^{1/2}_{th}=m_N^*$ is rather pronounced for
$s_0=(1000~\mbox{MeV})^{2}$ but it is only minor for
$s_0=(500~\mbox{MeV})^{2}$. Thus, we conclude that taking into
account  the energy dependence of the width might be quite
important for very broad resonances, when the width only slowly
decreases with energy. If the width drops rather rapidly with the
energy increase,   the quasiparticle approximation becomes
appropriate for calculation of thermodynamic quantities already at
not too high temperature. The baryon density dependence of the
ratio $R_{\Delta}$  is not so pronounced (especially for $s_0
=(500~\mbox{MeV})^2$), since in our parameterization the spectral
function depends on the density only through the value $m^*_{\rm
r}(n_B )$ and the choice of $s_{\rm th}$. For low temperatures
($T\lsim 50\div 100$~MeV), the $R_{\Delta}$ ratio becomes
significantly larger than unity. We also pay attention to the
shift of the reaction thresholds due to the dependence of the
width on $s$ and the threshold $s_{\rm th}$ on the density and
temperature. This point can be very important  for fitting of
particle momentum distributions.

To demonstrate the effect of the finite resonance width on
thermodynamic characteristics of the system,  we calculate the
energy density of the non-interacting resonances (however with
width). Then
 \be\label{efres}
E^{\rm f}_{\rm r}=N_{\rm r}\int_{0}^{\infty}\frac{4\pi p^2 d
p}{(2\pi)^3}\int_{0}^{\infty}\frac{ d s}{4\pi}\om \widetilde{A}_{
\rm r } f^{\rm f}_{\rm r }+(\mu_{\rm r}\rightarrow -\mu_{\rm r}),
 \ee
  where the degeneracy factor for $\Delta$ is $N_{\rm r}=16$.

In Figs. \ref{ratiosDe1} and  \ref{ratiosDe11},  we show the
ratios of the energy density for $\Delta$'s (with and without
width) to the total baryon energy density at $n_B =0$ and $n_B
=5~n_0$, respectively. The solid and dashed curves correspond to
calculational results for the width with $s_0
=(500~\mbox{MeV})^{2}$ and $s_0 =(1~\mbox{GeV})^{2}$. The
dash-dotted curve is computed within the quasiparticle
approximation. Calculations are performed for two values of the
threshold energies $s^{1/2}_{\rm th}=m_N^{*}$ (left panels) and
$s^{1/2}_{\rm th}=m_N^{*}+m_{\pi}$ (right panels).

As was expected, in all cases the quasiparticle result is much
closer to that for $s_0 =(500~\mbox{MeV})^{2}$ than for $s_0
=(1~\mbox{GeV})^{2}$. In the former case, the differences are
almost negligible. For $s_0 =(1~\mbox{GeV})^{2}$ the ratio remains
smaller than the quasiparticle result, except for low
temperatures. Differences in the ratios with and without taking
into account the width in all cases are not too noticeable. The
density dependence of the ratio proves to be pronounced even in
our model, although  the density dependence of the width is not
incorporated explicitly. Summarizing, for calculation of
thermodynamic characteristics, one may use the quasiparticle
approximation for baryons in the whole temperature interval under
consideration provided the high energetic tail of the resonance
width-function is not too long. If a resonance has a long
energetic tail, then the longer the width tail is, the more
suppressed the ratio of the resonance energy to the total energy,
as compared to the corresponding quasiparticle ratio.

\begin{figure}[thb]
 \hspace*{2mm}
\includegraphics[width=130mm,clip]{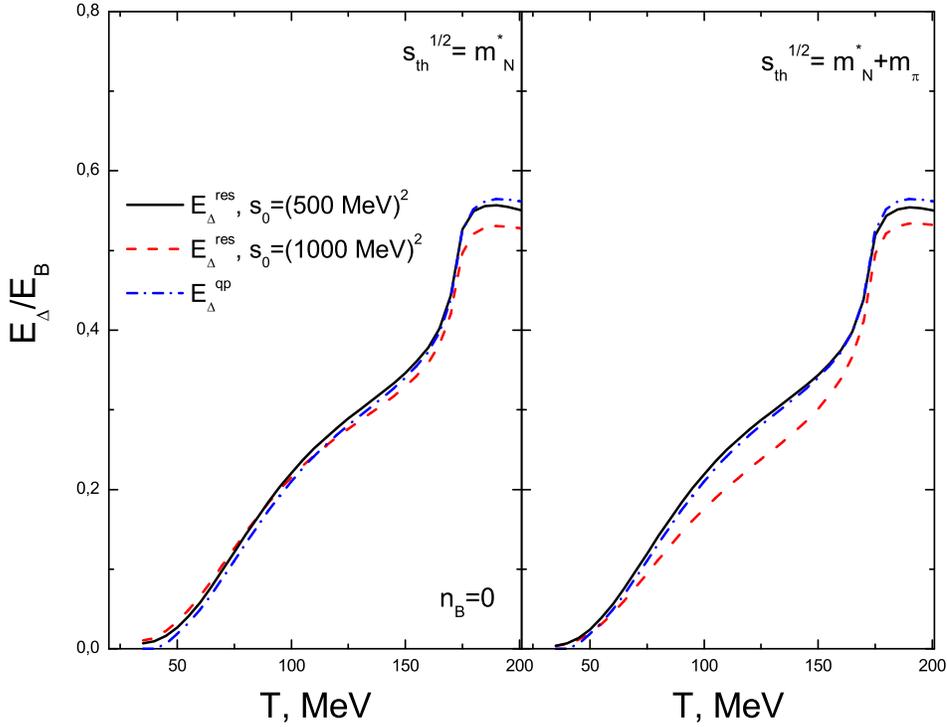}
\caption{  Ratio  of the $\Delta$ isobar energy density,
calculated with the inclusion of the width ($E_{\Delta}$), to the
total baryon energy density ($E_B$),  as a function of temperature
at $n_B =0$ for two values of the resonance threshold energy
$s_{th}^{1/2}$. The parameters of calculation are presented in the
figure. $E^{qp}_{\Delta}$ is calculated within the quasiparticle
approximation.
 }
 \label{ratiosDe1}
\end{figure}

\begin{figure}[thb]
 \hspace*{2mm}
\includegraphics[width=130mm,clip]{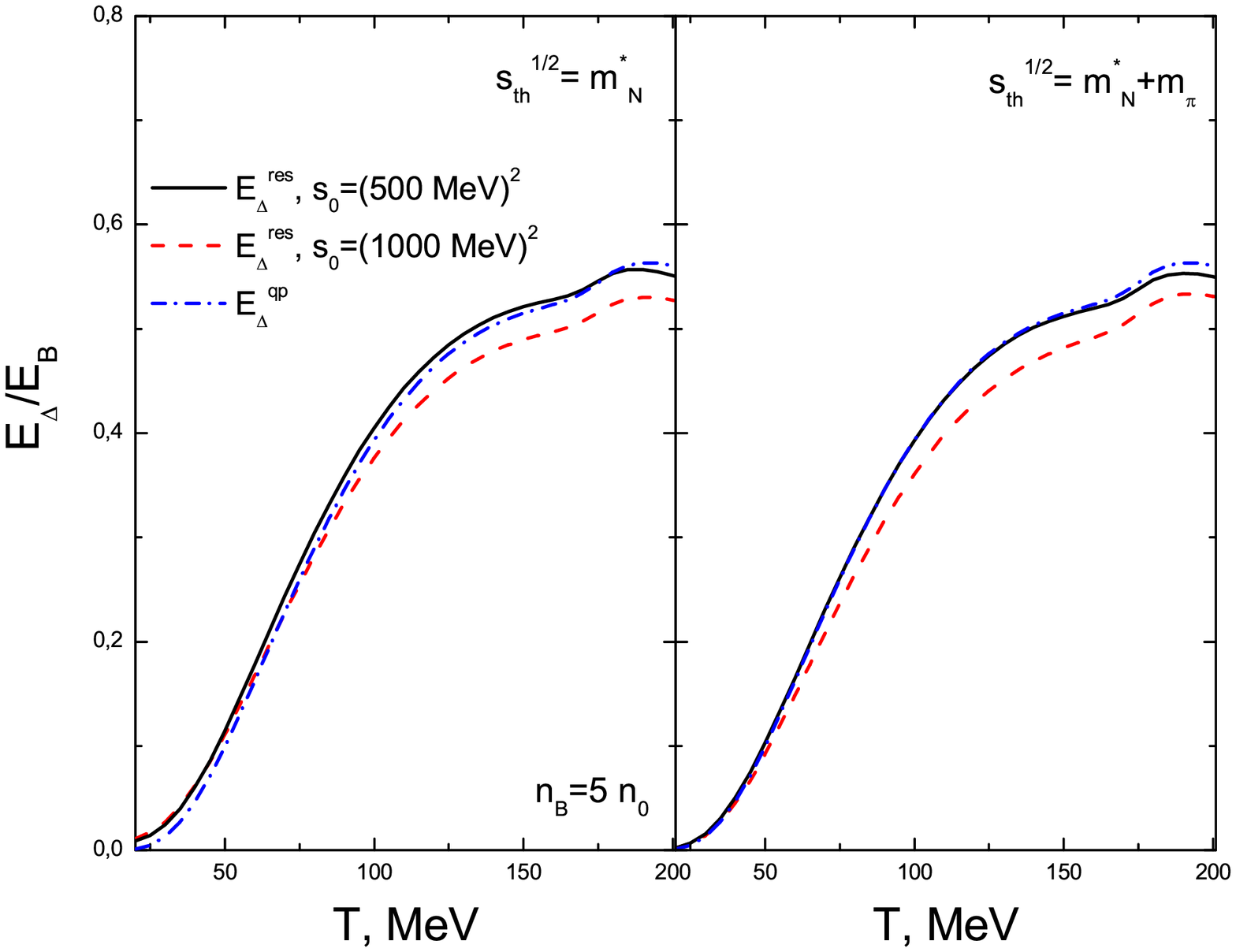}
\caption{ The same as in Fig. \ref{ratiosDe1} but for $n_B
=5n_0$.
 }
 \label{ratiosDe11}
\end{figure}

For charged bosons\footnote{As before, by the charge we mean any
conserved quantity like electric charge, strangeness, etc.} the
spectral function follows the sum-rule, cf. \cite{IKV3},
 \be
\int_{0}^{\infty}\frac{ d s}{2\pi}A^{\rm b}_{\rm r}=1.
 \ee
We again consider a  dilute matter assuming that the spectral
function depends only on the $s$-variable. Only in this case one
may consider a single spectral function for vector mesons, like
$\om$ and $\rho$, whereas in the general case one should introduce
transversal   and longitudinal components.

The Noether charged boson density (of given charge) is given by
 \be\label{Noethbos} n^{\rm b}_{\rm r}=(2s_{\rm
r}+1)\int_{0}^{\infty}\frac{4\pi p^2 d
p}{(2\pi)^3}\int_{0}^{\infty}\frac{ d \om}{2\pi} 2\om \ A^{\rm
b}_{\rm r} \ f^{\rm b}_{\rm r}, \quad f^{\rm b}_{\rm
r}=\frac{1}{e^{(\om -\mu_{\rm r}^{*})/T}-1}.
 \ee
For practical calculations it is convenient to present the
spectral function in the form:
 \be
 \label{fenomen1}
 A^{\rm b}_{\rm r}=\frac{\xi \
 [\Gamma_{\rm r}^{\rm b} (s)] }{(s -m^{*2}_{\rm r})^2 +[\Gamma_{\rm r}^{\rm b}
(s)]^2 /4 }+\xi\cdot 2\pi\delta (s-(m_{\rm r}^*)^2 ) \ \theta
(s_{\rm th}-(m_{\rm r}^*)^2 )
 \ee
 with $\xi =const$ introduced to fulfill the
sum-rule. Replacing $\widetilde{\Gamma}_{\rm r}
(s)=\frac{1}{2}\Gamma_{\rm r}^{\rm b} (s)$, we may use eq.
(\ref{widthf}) for $\widetilde{\Gamma}_{\rm r} (s)$ with $\alpha
=1/2$ for $s$ and $\alpha =3/2$ for the $p$ resonance.

\begin{figure}[thb]
 \hspace*{2mm}
\includegraphics[width=130mm,clip]{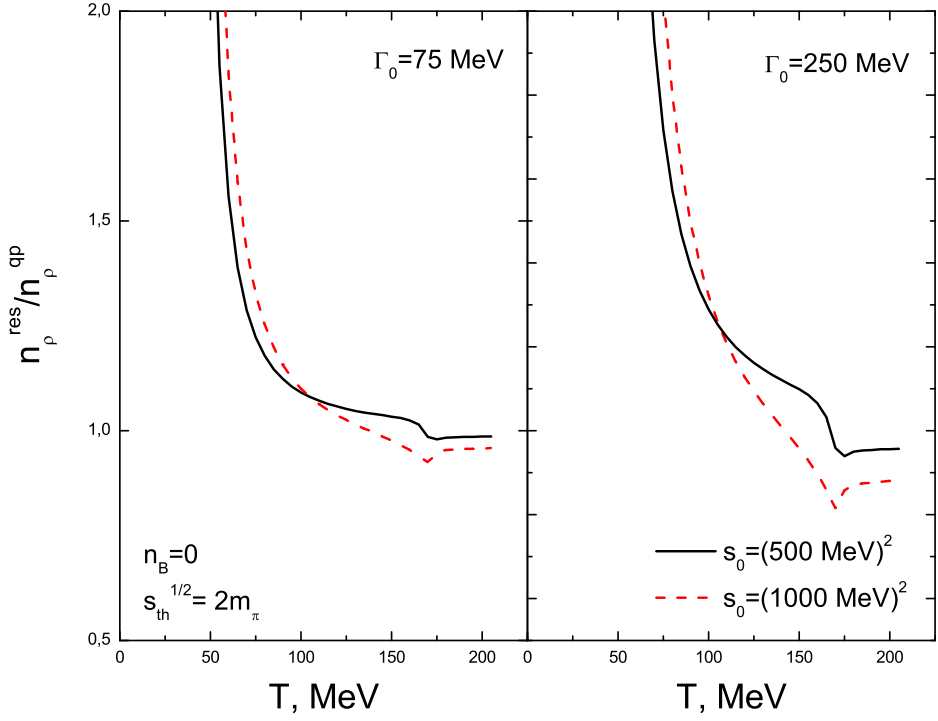}
\caption{ Ratio of the $\rho$ meson density calculated with the
inclusion of the width to the quasiparticle one, $R_{\rho}=n^{\rm
res}_{\rho}/n_{\rho}^{\rm qp}$, as a function of the temperature
at $n_B =0$. The parameters of calculation are presented in the
figure.
 }
 \label{ratiosD11r}
\end{figure}

\begin{figure}[thb]
 \hspace*{2mm}
\includegraphics[width=130mm,clip]{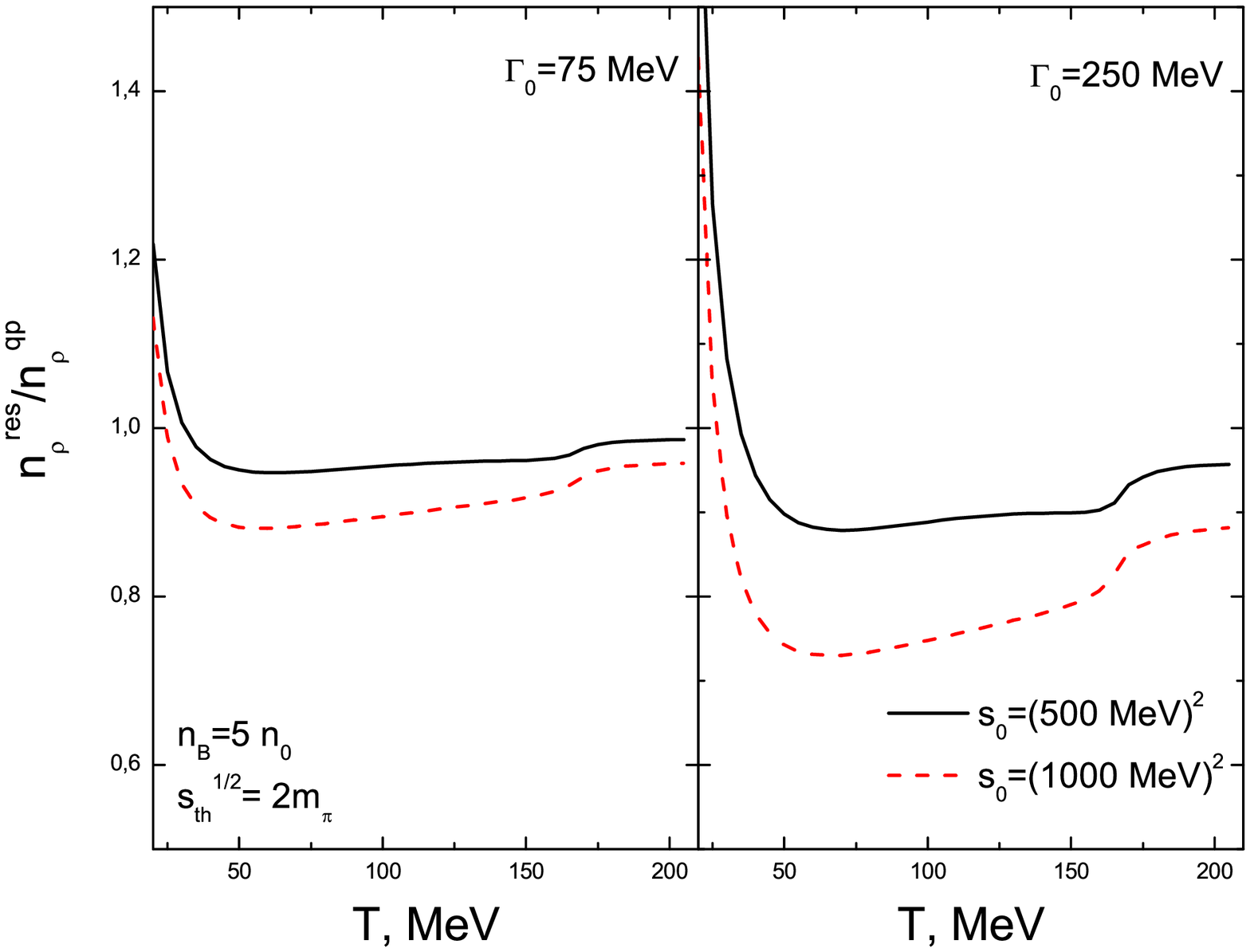}
\caption{ The same as in Fig. \ref{ratiosD11r} but for $n_B
=5~n_0$.
 }
 \label{ratiosD11rr}
\end{figure}

In Figs. \ref{ratiosD11r} and \ref{ratiosD11rr}, the ratio
$R_{\rho}=n_{\rho}/n^{\rm qp}_{\rho}$ of the $\rho^{+}$ density,
calculated following eq. (\ref{Noethbos}), to the quasiparticle
density is shown  as a function of the temperature. Here we take
into account the $p$-wave nature of the resonance and use two
values for the width:  a small width $\Gamma_0
=75~$MeV (left panels) and a very large one $\Gamma_0
=250~$MeV (right panels). We use $s_{\rm th}^{1/2}=2m_{\pi}$,
thus taking $m_1^* =m_2^* =m_{\pi}$. The results are presented for
$n_B =0$ and $n_B =5~n_0$ in Figs. \ref{ratiosD11r} and
\ref{ratiosD11rr}, respectively. These figures show only a
moderate dependence of the ratios $R_{\rho}$ on the value of the
width  at the resonance peak (on $\Gamma_0$).  At high energies
the energy dependence of the width is more pronounced (compare
solid and dash curves). In our model the density dependence of
$R_{\rho}$ reflects   the behavior of $m^{*}_{\rho}(n_B)$. For
$n_B =0$ the ratio $R_{\rho}$ becomes larger than unity for
$T\lsim 100$~MeV, whereas for $n_B =5~n_0$ $R_{\rho}$ is larger
than unity only for $T\lsim 20\div 50$~MeV. The ratio
$R_{\rho}(n_B =5~n_0)<R_{\rho}(n_B =0)$ since the resonance mass
$m^{*}_{\rho}(n_B)$ for $n_B =5~n_0$ is  closer to the threshold
value $2m_{\pi}$ than in the case of $n_B =0$.

 Slight bends of the curves at $T\sim 170$~MeV are
associated with dropping of the effective mass of $\rho$ below the
threshold. Then there arises a quasiparticle contribution to the
spectral function being added to the high-energy width term
(second term in (\ref{fenomen1})). In general, the behavior of
$R_{\rho}$ and $R_{\Delta}$ is similar. The quasiparticle
approximation is rather appropriate for $T\gsim 100$~MeV, provided
the widths have not too long high-energy tails (see curves with
$s_0^{1/2} =500$~MeV). Under these conditions our quasiparticle
SHMC model works well. If the resonance width has a very broad
high-energy tail (see curves with $s_0^{1/2} =1000$~MeV), the
deviation of the $R_{\rm r}$ ratio  (for the ${\rm r}$-resonance)
from unity is rather pronounced, even for $T\gsim 100$~MeV. Since
in this case $R_{\rm r}<1$, when  broad resonances are included,
there may appear a possibility to match the lattice data for
$T\gsim 170$~MeV  even without suppressing the coupling constants.
The latter procedure was used in \cite{KTV} to demonstrate a
possibility to fit the lattice results within our quasiparticle
SHMC model.

Broad hadronic resonances having a short lifetime are of a
particular interest for dynamics at the late stage of relativistic
heavy-ion collisions. As the system expands and cools, it will
hadronize and chemically freeze out (vanishing inelastic
collisions, no creation of new particles). After some period of
hadronic elastic  interactions, the system reaches the kinetic
freeze out stage, when all hadrons stop interacting at all
(vanishing even elastic collisions). After the stage of the
kinetic freeze out, particles overcoming remaining mean fields
free-stream towards the detectors, where measurements are
performed. The lifetimes of the $\rho$ meson and  $\Delta$-baryon
are $\tau_{ \rho} \simeq 1/\Gamma_{0\rho} \simeq 1.3~fm/c$ and
$\tau_{ \Delta} \simeq 1.7~fm/c$, respectively, being small with
respect to the lifetime of the expanding  system. So short-lived
resonances can decay and regenerate in scattering process all the
way through the kinetic freeze out. The regeneration process
depends on the hadronic cross sections of resonance daughters.
Thus the study of different resonances can provide an important
probe of the time evolution of the source from the beginning of
the chemical freeze out  to the kinetic one. In-medium effects can
modify properties of hadron quasiparticles and resonances during
this stage. In particular, it concerns finite widths and
energy-momentum relation evolving in time. Values of chemical
potentials and the temperature characterizing particle momentum
distributions become completely frozen at the kinetic freeze out.
From this  moment only the mean fields can survive and the hadron
effective masses further evolve towards the bare masses with which
particles reach detectors.

\begin{figure}[thb]
 \hspace*{4mm}
\includegraphics[width=130mm,clip]{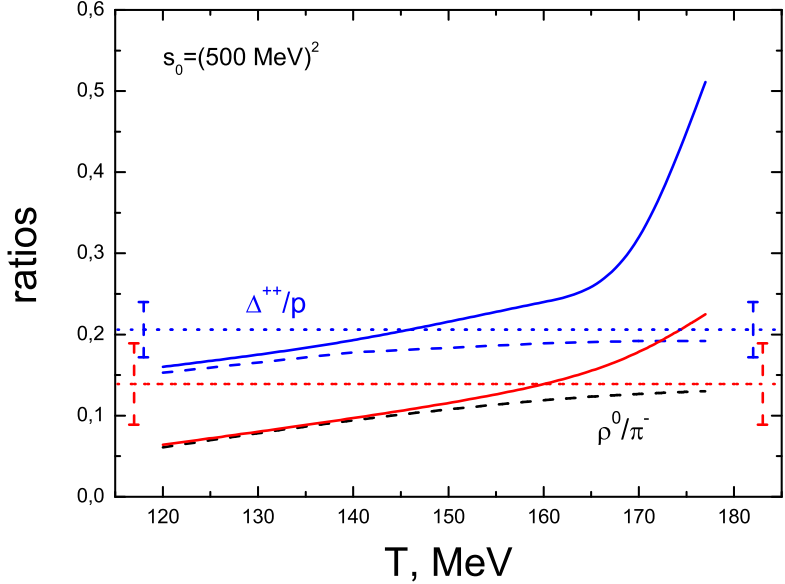}
\caption{ The resonance ratios as a function of temperature. The
solid lines are calculated within the SHMC model with accounting
for the resonance width according to eqs. (\ref{NoethB}),
(\ref{widthf1}). Parameters are the same as in Figs.
\ref{ratiosD1} and \ref{ratiosD11r} (right panels, solid lines).
The dashed lines are calculated for the IG model. These results
are obtained at $\mu_B=20$ MeV   for the RHIC energy. Experimental
data~\cite{dAuStar08} for central $d+Au$ collisions are plotted by straight
dotted lines with error bars.
}
 \label{rho_D_ratio}
\end{figure}

In Fig. \ref{rho_D_ratio} the ratios of yields of resonances to
the yields of stable particles with similar quark content, the
$\rho^0/\pi^-$ and $\Delta^{++}/p$ ratios, are presented as a
function of temperature. For all four species considered, our
calculations take into account the feed-down from higher
resonances\footnote{Those hadrons from the particle data table
which originally were not included into the SHMC model set are
treated here as the ideal gas of resonances with vacuum masses and
vanishing widths.} $n^{feed}_i=n_i+\sum_r n_r \Gamma^{r\to
i}/\Gamma_r$. Being in an agreement with experiment, the
statistical analysis of {\em stable} hadron ratios at the chemical
freeze-out \cite{ABMS05} shows that these ratios are getting
almost energy-independent at $\sqrt {s_{NN}}\gsim 100$ GeV. This
statement seems to be valid also for {\em resonances} provided
they are treated within the quasiparticle approximation, as
follows from the weak $T$ dependence of the ideal gas (IG) model
\footnote{It is a quasiparticle model with the vacuum masses for
all hadrons. Mesons with masses $m_i\leq 1.6$ GeV and baryons with
$m_i\leq 2.2$ GeV are included.} results (see dashed lines in Fig.
\ref{rho_D_ratio}).  As is seen in Fig.~\ref{rho_D_ratio}, the resonance
ratios calculated in the SHMC and IG models almost coincide with each
other till temperature about 140 and 155 MeV for the $\Delta^{++}/p$ and
$\rho^0/\pi^-$ ratios, respectively,  and then the
difference between them increases with $T$. The growth of the
resonance yield in the SHMC model is mainly due to the dropping of
the effective mass at high temperatures. Dependence of the
resonance abundance on the value of the width and other parameters
(at the same temperature) is rather moderate (compare the solid
and dashed curves in Figs. \ref{ratiosD1} and \ref{ratiosD11r}).

Earlier, the resonance statistical treatment has been considered
in refs. \cite{BGK04} and \cite{BMRS03}. Both models are ideal gas
ones  but in the hadronization statistical model \cite{BMRS03} the
vacuum (energy independent) resonance widths are included. At the
same values $T,\mu_B$ these models provide quite close results but
if one refers to the particular collision energy, their results
differ due to making use of different approximations for the
freeze out $T$ and $\mu_B$ as functions of the collision energy
(see Fig. \ref{Trad1}). In the hadronization statistical
model~\cite{BMRS03}, predictions were made only for the SPS
energies.

Very recently the short-lived resonances $\Delta^{++}$ and
$\rho^0$ have been measured by the STAR Collaboration at RHIC in
$d+Au$ collisions~\cite{dAuStar08}. The observed $\rho^0/\pi^-$
and $\Delta^{++}/p$ ratios practically do not depend on the
charged particle number at the mid rapidity, $dN_{ch}/d\eta$ ({\em
i.e.} on centrality), and for central ($20\%$ centrality)
collisions are $0.139\pm 0.050$ and $0.206\pm 0.034$,
respectively~\cite{dAuStar08}. As is seen in
Fig.~\ref{rho_D_ratio}, calculated results for the SHMC model
cross the experimental line at $T\sim 145$ and $\sim 160$ MeV
\footnote{The temperature concept should be used with care for
$d+Au$ collisions. In experiments, temperature is usually
associated with the inverse slope of transverse mass
distributions.} for $\Delta^{++}/p$ and $\rho^0/\pi^-$,
respectively. However, large experimental error bars do not allow
to make preference to any specific model, though the IG model
predictions are slightly but regularly below that for the SHMC
model. One should remind that at the RHIC energy the chemical
freeze out temperature is estimated as 177
MeV~\cite{BMRS03,dAuStar08} and kinetic one is about 120 MeV. The
$\rho^0/\pi^-$ ratio was also measured in peripheral $Au+Au$(200
GeV) collisions~\cite{AuAuStar04} to be as large as
$\rho^0/\pi^-=0.169\pm 0.003(stat)\pm 0.027(syst)$.
 Note that the existence of the difference between the
experimental value of $\rho^0/\pi^-$ ratio and the statistical
model result has been considered as a problem in ref.
\cite{BMRS03}. Following their estimate, $\rho^0/\pi^-$ ratio is
about $4\cdot 10^{-4}$ at the kinetic freeze out temperature  120
MeV, and it is  0.11 at the chemical freeze out $T=177$ MeV. Our
result ($\sim 0.06$ at $T=120$ MeV) strongly deviates from the
statistical estimate~\cite{BMRS03} at the kinetic freeze out. At
the chemical freeze out temperature, $T=177$ MeV, predictions of
the $\rho^0/\pi^-$ ratio  practically coincide in all ideal gas
based models remaining lower than the experimental ratio, whereas
our SHMC model is able to reproduce experiment at $T\sim 165\div
170$~MeV.  It is also of interest that the measured resonances
have a mass shift about 50-70 MeV ~\cite{dAuStar08} which is
consistent with the SHMC model results at $T\sim 160$ MeV as
presented in Fig. \ref{KTV}. Thus, one may infer that due to a
possible decay, regeneration and rescattering, the broad
resonances, if they are treated within the statistical picture
with taken into account their in-medium mass shift, freeze at a
somewhat lower temperature than that at the chemical freeze out
but this temperature is significantly higher than the value of
$T\approx 120$~MeV, characterizing the kinetic freeze out
\cite{ShB03}. A dynamical consideration of the freeze out is
needed to draw more definite conclusions.

\section{Concluding remarks}\label{Concluding}
In this paper we made attempts to find several improvements of the
SHMC model of \cite{KTV}.

In \cite{KTV} the boson excitation effects were assumed to be
small. Therefore their contribution to thermodynamic
characteristics was calculated using perturbation theory in the
fields of boson excitations. Thereby, boson excitation
contributions in the equations of motion  were dropped.  The
approximation made needs justification. Here we incorporated the
boson excitation terms in the equations of motion and then
calculated thermodynamic quantities (including boson excitation
parts). Corrections to the effective masses of the $\sigma$, and
$\om$-$\rho$-nucleon excitations turn to be minor for $T\lsim
100\div 120$~MeV and grow with the temperature increase.
Corrections to thermodynamic characteristics, like total pressure,
energy, entropy {\em etc.}, remain moderate even at higher
temperatures. Qualitatively, one may conclude that all results of
\cite{KTV} remained unchanged. With boson excitation terms
incorporated into the equations of motion, our quasiparticle SHMC
model fulfills exactly the thermodynamic consistency conditions.

Then we discussed possible effects of the resonance widths. We
assumed vacuum {\em but energy-dependent}  widths of resonances.
Under this assumption the
 width effects included do not change  qualitative behavior of
the system. Nevertheless, one should note that in dense and/or hot
matter particle widths may acquire essential density- and
temperature-dependent contributions that may significantly affect
properties of the system,  e.g., see \cite{V04}, where for the
case of hot baryon-less system it is shown that width effects may
completely smear fermion distributions.  We illustrated that the
estimated yields of short-lived resonances to be important for the
late stage of relativistic heavy-ion collisions can be described
sufficiently well if  one treats those resonances in the framework
of our SHMC model with the vacuum widths at the freeze out.

In the paper, the $\sigma$ variable was  considered as an order
parameter. Then the effective masses of $\sigma^{'}$, $\om^{'}$,
$\rho^{'}$ excitations and the effective masses of nucleons have a
similar behavior as a function of the baryon density and
temperature. The effective masses remain rather  flat functions of
the temperature and sharply drop to zero only in the vicinity of
$T_c$. In the large temperature interval mentioned, the effective
masses as a function of the density first decrease with  the
density increase and then, at a very high density, begin to grow,
cf. \cite{KTV}.

In Appendix A we calculated what modifications of our model could
be, if  the $\sigma$-$\om$ and $\rho$ fields were treated on equal
footing. In the latter case, the density-temperature behavior
of the $\sigma^{'}$ excitation mass essentially differs from that
of the $\om^{'}$-$\rho^{'}$ excitation masses, namely, the
$\sigma^{'}$ excitation mass drops to zero at $n_c \simeq 4.2~n_0$
and only slightly depends on $T$. Thus, if this model was applied
to analyze heavy ion collisions, we would face with a problem of
Bose condensation of $\sigma^{'}$ excitations for $n_B >n_c \sim
4\div 5~n_0$ in a broad temperature interval. Another unpleasant
feature of a model version like this is a drastic difference in
the density behavior of the $\sigma^{'}$ effective mass and that
of $\om^{'}$-$\rho^{'}$ excitations.  These are additional
arguments in favor of the treatment of the $\sigma$ variable as an
order parameter (as we did earlier in ref. \cite{KTV} and here in
the paper body).

Then in Appendix B an attempt was made to incorporate the
nucleon-nucleon hole and nucleon-antinucleon loop effects in our
model. The loop terms mentioned were calculated within the
perturbative approach. If these terms are taken into account, the
$\sigma^{'}$ excitation effective mass exhibits rather unrealistic
behavior. Thus, the model completely loses its attractiveness if
baryon loops are included (at least within the perturbative
approach). We have checked that this unpleasant feature is a
common feature of many RMF models, including the original Walecka
model  (previously authors of ref. \cite{FPS} arrived at a
similar conclusion considering vacuum loop corrections in the
Walecka model). In the Fermi liquid approach to diminish
contribution of the fermion loop terms one incorporates the
vertices corrected by a short-range baryon-baryon interaction
introduced with the help of the Landau-Migdal parameters, cf.
\cite{MSTV90,V93,KV03}. In the framework of our SHMC model
short-range correlation effects are simulated by the $\om$,
$\sigma$, $\rho$ exchanges and are non-local. Thereby, we do not
include these effects in the present work. Summarizing, either
higher order fluctuation effects should be included in all orders,
that is a complicated problem, or they should be skipped within
RMF based models. So we skipped these terms in our truncated
scheme.

Concluding, the SHMC model introduced in ref. \cite{KTV} can be
considered as a reasonable model for  application to the
description of hadronic matter  in a broad baryon density-temperature 
range, provided higher order fluctuations of fermion
fields are not included.

 \vspace*{5mm} {\bf Acknowledgements}
\vspace*{5mm}

We are very grateful to  A.~Andronic, Yu.B. Ivanov,
E.E.~Kolomeitsev, K. Redlich, and V.V. Skokov for numerous
illuminating discussions, valuable remarks, and constructive
criticism.  This work was supported in part by the Deutsche
Forschungsgemeinschaft (DFG project 436 RUS 113/558/0-3), and the
Russian Foundation for Basic Research (RFBR grants 06-02-04001 and
08-02-01003).

{\bf{Appendix A. Treatment of the $\sigma$ field as an
independent variable.}}\label{ind}
 As noted in the paper,  we continue to suppress
contributions of baryon loops. If the $\sigma$ and  $\om_0$ fields
are treated on equal footing, i.e. as independent variables
("ind.var."), one can determine the effective $\sigma^{'}$-mass
using partial derivatives of the pressure,
 \be
 \label{micro}(m_{\sigma}^{\rm part*})^2_{\rm ind.
var.}=-\frac{{\partial^2 \sum_{m\in\{m\}}P_m^{\rm MF}[f
,\omega_0]}}{{\partial f^2}}\left(\frac{d f}{d \sigma}\right)^2 ,
 \ee
 rather than full derivatives.

It is convenient to present
 \be \label{Pspa11}  &&\frac{{d^2
\sum_{m\in\{m\}}P_m^{\rm MF}[f ,\omega_0 (f )]}}{{d f^2}} =
\frac{{\prt^2 \sum_{m\in\{m\}}P_m^{\rm MF}[f ,\omega_0 ]}}{{\prt
f^2}}+\delta_{\sigma},\nonumber\\ &&\delta_{\sigma} \simeq
\frac{\partial^2 \sum_{m\in\{m\}}P_m^{\rm MF}}{\partial \om_0^2 }
\left(\frac{\partial \om_0}{\partial f}\right)^2 +2
\frac{\partial^2\sum_{m\in\{m\}} P_m^{\rm MF}}{\partial \om_0
\partial f}\frac{\partial \om_0}{\partial f}.
 \ee
  Taking derivatives  in (\ref{micro}) and
(\ref{Pspa11})  we use the exact equation of motion for $f$ and
$\om_0$, and  suppress the boson excitation and baryon-loop
contributions to the pressure in the final expression.

Taking partial derivatives we get
 \be\label{mfmes} -\frac{{\prt^2
 \sum_{m\in\{m\}}P_m^{\rm MF}[f ,\omega_0
]}}{{\prt f^2}}\simeq
\frac{m^4_N}{C_{\sigma}^2}+U^{\prime\prime}_f -\om_0^2 m_{\om}^2 ,
 \ee
 and $\delta_{\sigma}=-\om_0^2 m_{\om}^2$, where
$\eta_{\sigma}=1$ was used.

If  the  effective $\sigma^{'}$-excitation mass squared is
determined using the Hamiltonian  we obtain
 \be\label{mfmes0} \left<\frac{{\prt^2 H[f
,\omega_0 ]}}{{\prt f^2}}\right>=
\frac{m^4_N}{C_{\sigma}^2}+U^{\prime\prime}_f -\om_0^2 m_{\om}^2 +
\om_0\frac{\prt^2\chi_\om}{\prt f^2}\sum_{b\in\{b\}} g_{\om b}t_b
n_b ,
 \ee
 that differs from (\ref{mfmes}) by the last term.

\begin{figure}[thb]
 \hspace*{2mm}
\includegraphics[width=130mm,clip]{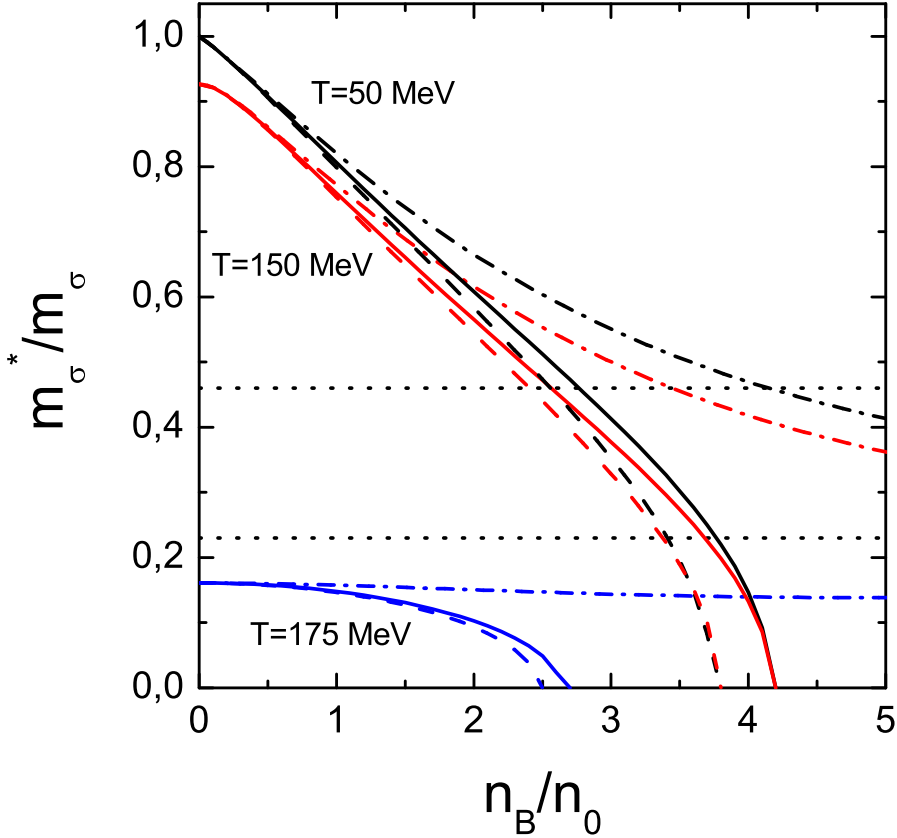}
\caption{ The effective $\sigma^{'}$ excitation mass as a function
of density calculated using eqs. (\ref{micro}) (solid curves),
(\ref{mfmes0}) (dash lines), and (\ref{spa3}) (dash-dotted curves)
for different temperatures. Straight dotted lines correspond to
$m_\sigma=2m_\pi$ and $m_\sigma=m_\pi$.
 }
 \label{MsgrPD}
\end{figure}

In Fig.\ref{MsgrPD},  we show  the ratio of the effective-to-bare
masses of the $\sigma^{'}$  excitation as a function of the baryon
density at different values of the temperature for two possible
treatments of the $\sigma$-field, as an order parameter, see
(\ref{spa3}), and as an independent variable. In the latter case
two expressions, (\ref{micro}) and (\ref{mfmes0}),  were used to
calculate the value $m_{\sigma}^{\rm part *}$. The solid lines
 are evaluated following eq.(\ref{micro}) for $T=50, \ 150$ and 
 $175$ MeV (from the top to the bottom); and the dashed
curves, using (\ref{mfmes0}). In both the cases the effective
$\sigma^{'}$ excitation mass, calculated following (\ref{micro}),
drops to zero at $n_B \approx 4.2 n_0$ (for $T=50$~MeV) and for
$n_B \approx 3.8~n_0$ (for $T=150$~MeV), respectively, which
demonstrates a moderate temperature dependence of the effective
mass up to $T\sim 150~$MeV. For higher temperatures the critical
density significantly decreases. In all the cases, the differences
between calculations following (\ref{micro}) and (\ref{mfmes0})
are minor; however, they significantly deviate from calculational
results following eq. (\ref{spa3}) (dash-dotted curves). The
behavior of the effective $\sigma^{'}$ excitation mass calculated
using (\ref{micro}) and (\ref{mfmes0}) is in contrast with the
behavior of the $\om^{'}$-$\rho^{'}$-$N$ effective masses which do
not reach zero for all densities (see dash-dotted curves). Since
one of our main goals was to construct a model based on the idea
of a rather similar behavior for all
$\sigma^{'}$-$\om^{'}$-$\rho^{'}$-$N$ excitation masses, we refuse
considering the $\sigma$ field, as an independent variable, i.e.,
we refuse to consider it on equal footing with the $\om$ and
$\rho$ fields. Therefore, instead of either eq. (\ref{micro}) or
eq. (\ref{mfmes0}), here we use eq. (\ref{spa3}) treating the
$\sigma$ field as an order parameter.

{\bf{Appendix B. Inclusion of baryon loops.}}\label{bario}

In the general case, the $\sigma^{'}$  excitation mass is given by
eq. (\ref{Pspa1}), provided  the $\sigma$ field is treated as an
order parameter. We will continue to keep only quadratic terms in
fluctuating boson fields. Thus, we drop the boson excitation term
in the pressure in expression (\ref{Pspa1}) but we will keep the
baryon term. So in contrast with (\ref{spa3}), we use the
expression
 \be
(m_{\sigma}^{\rm part*})^2 \simeq &-& \left[\frac{{d^2
\sum_{m\in\{m\}}P_m^{\rm MF} [f ,\omega_0 (f )]}}{{d f^2}} \right.
\nonumber \\ &+& \left. \frac{{d^2 \sum_{b\in\{b\}}P_b [f
,\omega_0 (f )]}}{{d f^2}} \right] \left(\frac{d f}{d
\sigma}\right)^2 . \label{Pspa2}
 \ee
An important difference between (\ref{Pspa2}) and (\ref{spa2})
(the latter expression yields the same result as  (\ref{spa3})) is
that derivatives of the Hamiltonian are taken at fixed $\Psi_B$,
whereas the total pressure $P$ depends on the baryon occupations,
which should be varied. Thereby, (\ref{Pspa2}) includes extra
contributions from the baryon-baryon hole and baryon-antibaryon
excitations.

Let us now calculate an additional, purely baryon contribution to
the $\sigma^{'}$-excitation mass incorporated in (\ref{Pspa2}).
Using that
 \be\frac{\partial
f_b}{\partial f}=\frac{\partial f_b}{\partial
\om_b}\left[\frac{\partial \om_b}{\partial f}+g_{\om b}\om_0
\frac{\partial \chi_{\om}}{\partial f}\right],\quad
\frac{\om_b}{p}\frac{\partial f_b}{\partial p} =\frac{\partial
f_b}{\partial \om_b},
 \ee
 and (\ref{cons1}), with the help of partial differentiations
 we find
 \be
 \label{deltab}
&&\delta_B (m_{\sigma}^{\rm part*})^2 =-\left(\frac{d f}{d
\sigma}\right)^{2}\sum_{b\in\{b\}}\frac{{\prt^2 P_b [f ,\omega_0
]}}{{\prt f^2}},\nonumber\\ &&-\sum_{b\in\{b\}}\frac{{\prt^2 P_b
[f ,\omega_0 ]}}{{\prt f^2}}=m_N^2\sum_{b\in\{b\}}x_{\sigma b}^2
\left[\frac{2 n_{sb}}{m_b^{*}}- L_b \right]
+\sum_{b\in\{b\}}g_{\om b}\om_0 \ t_b \ n_b \ \frac{\partial^2
\chi_{\om}}{\partial f^2}\nonumber\\
 &&-\sum_{b\in\{b\}}
\left(g_{\omega b}\om_0 \frac{\partial\chi_\omega}{\partial
f}\right)^2B_b +2m_N\sum_bx_{\sigma b}g_{\omega
b}\om_0\frac{\partial\chi_\omega}{\partial f} \widetilde{L}_b .
 \ee
We also use the equations of motion and the relation
$\left(\frac{\partial m_b^{*}}{\partial f}\right)=-x_{\sigma
b}m_N$.

 Here the quantities
 \be B_b =L_b +\frac{ n_{sb}}{m_b^{*}},
  \ee
and
 \be L_b = N_b \ \int_{0}^{\infty}\frac{dp}{2\pi^2} \ \om_b
 \ f_b
 \ee
are the baryon-baryon hole and baryon-antibaryon loop-terms taken
at zero incoming energy and momentum.

\begin{figure}[thb]
\begin{center}
\includegraphics[width=130mm,clip]{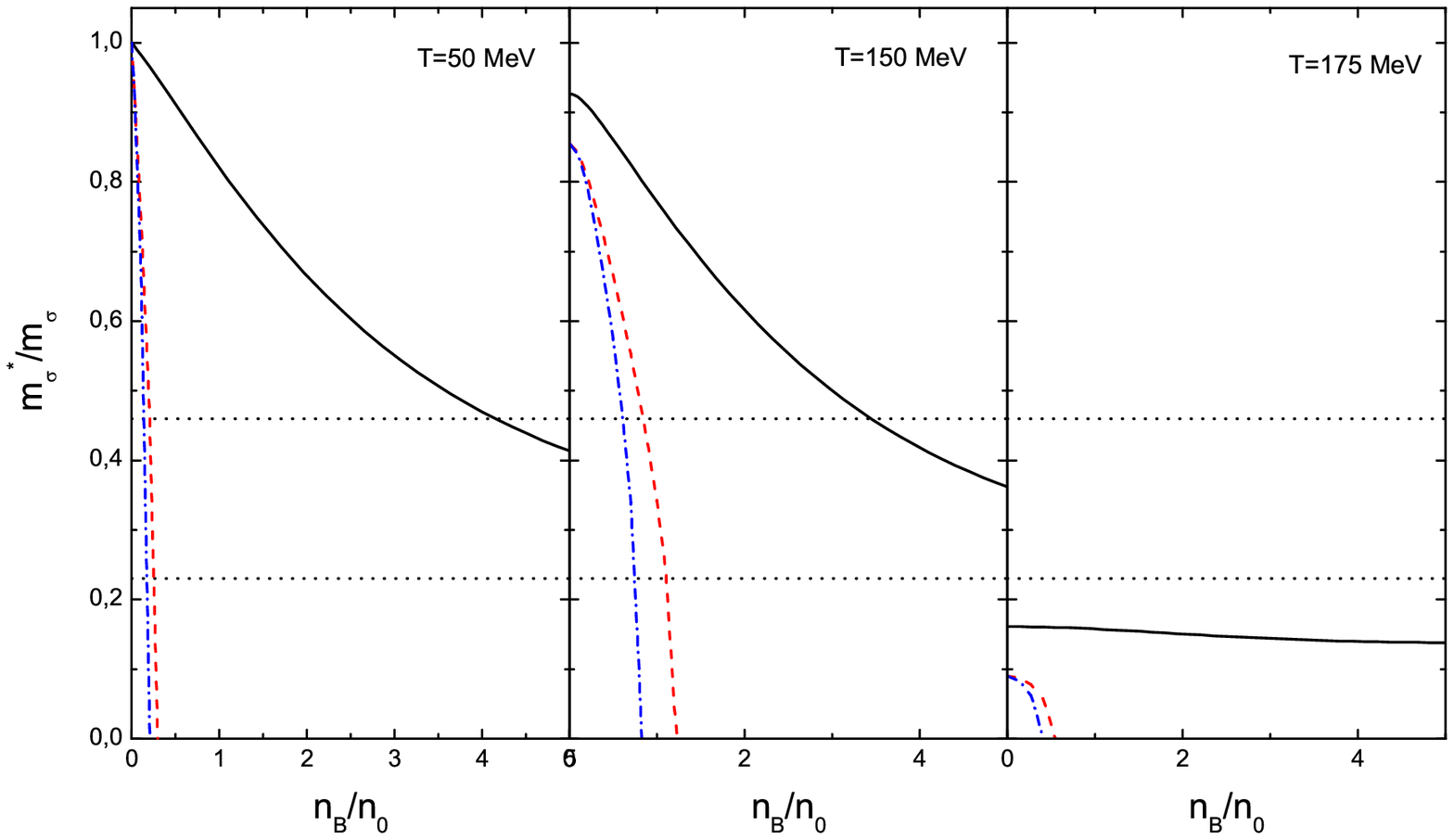}
\includegraphics[width=130mm,clip]{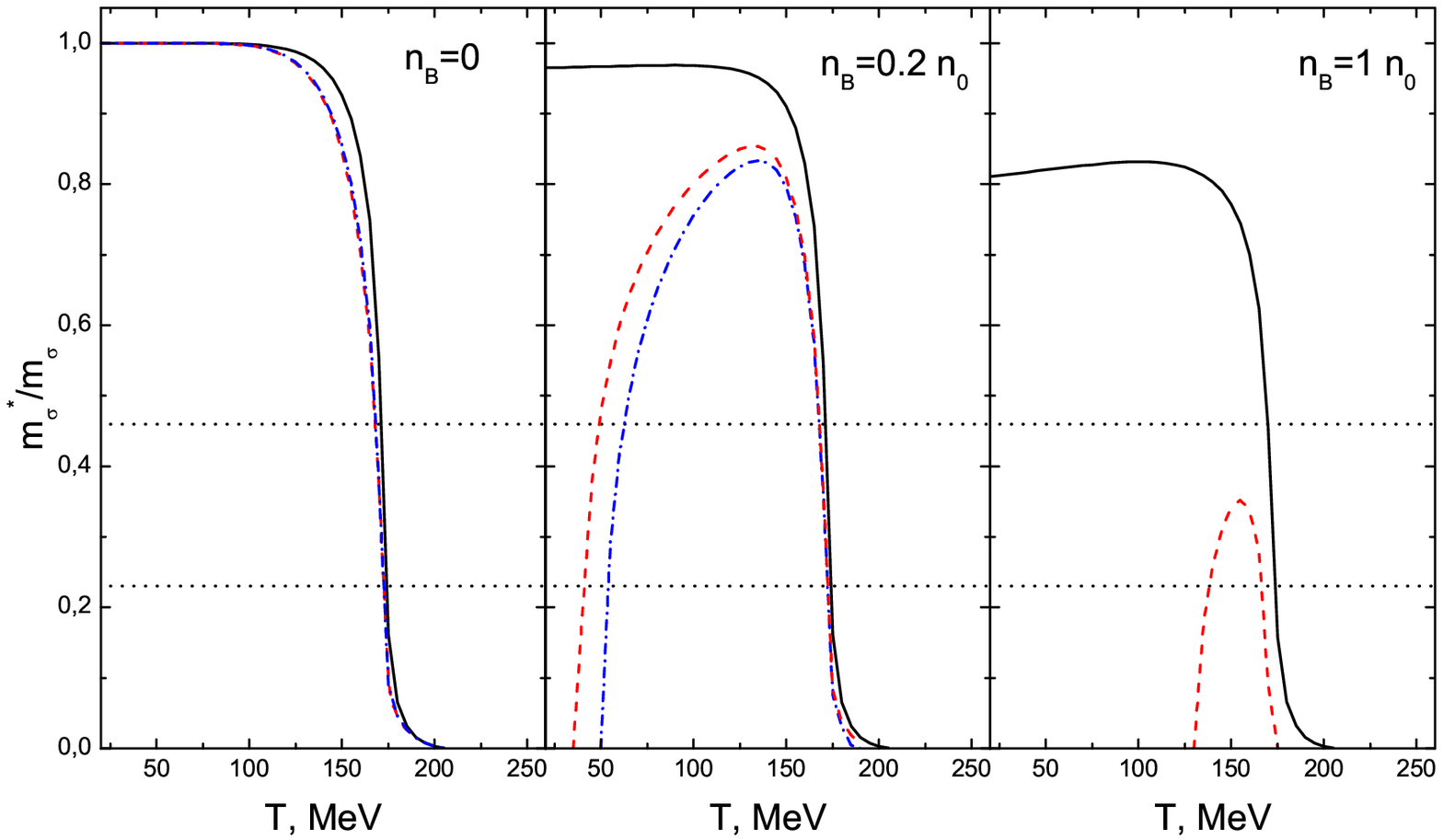}
 \caption{ The ratio
of the effective $\sigma'$ excitation mass to the bare mass as a
function of density (top) and  temperature (bottom) calculated by
means of eqs. (\ref{spa3}), (solid curves), eq. (\ref{micro2})
(dashed curves) and eq. (\ref{spa1}) (dash-dotted lines), see
(\ref{Pspa2}). The straight dotted lines correspond to the level
of $m_\sigma=2m_\pi$ and $m_\sigma=m_\pi$.
 }
 \label{MsgrPD1}
  \end{center}
\end{figure}

We also introduce a similar quantity
  \be
\widetilde{L}_b = N_b \ \int_{0}^{\infty}\frac{dp}{2\pi^2} m_{b}^*
\ t_b \ f_b .
 \ee

Then we calculate an additional contribution
 \be
\label{addit} &&\delta =-\frac{{d^2 P [f ,\omega_0 (f )]}}{{d
f^2}}+\frac{{\partial^2 P [f ,\omega_0 ]}}{{\partial
f^2}}\nonumber\\&&= (m_{\om}^{part*.0})^2\left(\frac{\partial
\om_0}{\partial f}\right)^2+2\left[m_\om^2\om_0\frac{\partial
{\Phi_\om}^2}{\partial f}-\sum_{b\in\{b\}}g_{\om b} \ t_b \ n_b
\frac{\partial \chi_\omega}{\partial
f}\right.\nonumber\\&&\left.+\chi_\om
\om_0\frac{\partial\chi_\om}{\partial f}\sum_{b\in\{b\}} g_{\om
b}^2\,B_b-m_N\chi_\omega \sum_{b\in\{b\}} g_{\om b}x_{\sigma
b}\widetilde{L}_b\,\right]\frac{\partial \omega_0}{\partial f} ,
 \ee
that distinguishes full and partial derivative terms. Here
 \be
 \label{omf} &&\frac{\partial \om_{0} }{\partial f}\left[
1+ \sum_{b\in\{b\}}\frac{g_{\om b}^2}{m_{\om}^2
\eta_{\om}}B_b\right]= \sum_{b\in\{b\}}\frac{g_{\om b}}{m_{\om}^2
\eta_{\om}\chi_{\om}} \left[x_{\sigma b}m_N \widetilde{L}_b -B_b
g_{\om b}\frac{\partial \chi_{\om} }{\partial f}\,\om_0 \right]
\nonumber\\ &+& \sum_{b\in\{b\}}\frac{g_{\om b} \ t_b \
n_b}{m_{\om}^2}\frac{\partial [\eta_{\om}\chi_{\om}]^{-1}
}{\partial f}
 \ee
with $(m_{\om}^{part*.0})^2= {\partial^2 P^{\rm MF}}/{\partial
\om_0^2 }$.

In \cite{KTV},  the baryon excitation loop $L_b$ contributions
were suppressed. On the other hand, one may check that $L_b >
n_{sb}/m_{b}^{*}$, $L_b >\widetilde{L}_b$. Therefore, in
\cite{KTV} we disregarded the term $\delta_B (m_{\sigma}^{\rm
part*})^2$ as the whole, as well as the terms $\propto B_b$ and
$\widetilde{L}_b$ in (\ref{addit}) and (\ref{omf}). Thus in
\cite{KTV} eq. (\ref{spa3}) is actually used, being compatible
with (\ref{spa2}).

Incorporating the loop terms mentioned, one should use the
following expression
 \be
\label{spa1} {{(m_{\sigma}^{\rm part*})^2
=\left[-\!\!\!\!\sum_{m\in\{m\}}\frac{{{\prt}^2 P_m^{\rm MF}[f
,\omega_0 ]}}{{{\prt} f^2}} + \delta_B (m_{\sigma}^{\rm part*})^2
+\delta\right]\! \left(\frac{d f}{d \sigma}\right)^2 ,
 }}\ee
 where partial contributions are given by
(\ref{mfmes}), (\ref{deltab}) and (\ref{addit}).

If  $\sigma$ and  $\om_0$  are treated on equal footing, i.e. as
independent variables ("ind.var."), one should use
 \be
 \label{micro2}(m_{\sigma}^{\rm part*})^2_{\rm ind.
var.}=-\frac{{\partial^2 P [\sigma ,\omega_0 ]}}{{\partial
\sigma^2}}
 \simeq -\frac{{\prt^2 P [f ,\omega_0 ]}}{{\prt
f^2}}\left(\frac{d f}{d \sigma}\right)^2.
 \ee

The ratio of effective-to-bare masses of the
$\sigma^{'}$-excitation  is shown in Fig. \ref{MsgrPD1} as a
function of the baryon density for three values of temperatures
(top panel) and as a function of the temperature for three values
of the baryon density $n_B$ (bottom panel) calculated by means of
eq. (\ref{spa1}), i.e., following (\ref{Pspa2}). Results are
compared with (\ref{micro2}) and  (\ref{spa3}). As we can see, the
inclusion of the baryon loop terms (performed within the
perturbation theory)  completely destroys all achievements of our
SHMC model (besides the case $n_B=0$ when these loop terms are
suppressed).

Thus, at the moment we can see no simple way to generalize the
mean field based the SHMC model which would include
baryon-antibaryon and baryon -- baryon hole fluctuations.

\end{document}